\documentclass[english,aps, preprint]{revtex4-1}
\usepackage[T1]{fontenc}
\usepackage{textcomp}
\usepackage[utf8]{inputenc}
\usepackage{color}
\usepackage{babel}
\usepackage{amsmath}
\usepackage{amssymb}
\usepackage{graphicx}
\usepackage[pdfusetitle,
 bookmarks=true,bookmarksnumbered=false,bookmarksopen=false,
 breaklinks=false,pdfborder={0 0 1},backref=false,colorlinks=true]
 {hyperref}
\begin{document}
\title{Stability and squeezing of the three-photon degenerate parametric
down-conversion}
\author{Vinícius V. Seco}
\author{Alencar J. de Faria}
\email{alencar.faria@unifal-mg.edu.br}

\affiliation{Instituto de Ciência e Tecnologia, Universidade Federal de Alfenas,
CEP 37715-400, Poços de Caldas, MG, Brazil}
\begin{abstract}
Nonlinear multiphoton conversion processes are theoretically and experimentally
challenging systems in quantum optics and related areas. These phenomena
bring remarkable quantum properties and can be of fundamental relevance
in quantum information and computation applications. Based on recent
experimental advances and the possible uses in generalized three-photon
squeezing, we investigate degenerate three-photon parametric down-conversion
generated inside a cavity fed by a classical pump. In addition, we
consider that the process is stimulated by a coherent driving mode
and subjected to dissipation by spontaneous emission of the cavity.
A treatment using phase-space methods and stochastic differential
equations is applied and the stability conditions of its steady states
are found. We find that the system does not exhibit multistability,
instead it has a limited parameter region leading to a single stable
steady branch, alongside multiple unstable branches. Focusing on the
stable steady states and linearizing the dynamics around them, we
calculate the spectral densities of the quadrature variables. We identify
and characterize the squeezing of three-photon down-conversion. Moreover,
since the quasi-probability differential equations of the higher-order
multiphoton processes have higher-order derivatives, hindering their
mapping into stochastic differential equations, we make use of a positive
Wigner function approach to calculate two-dimensional spectral densities
associated with three-time correlation functions, with the goal of
studying non-Gaussian properties of the system.
\end{abstract}
\maketitle

\section{Introduction}

Optical parametric conversion processes are a key resource for implementing
quantum states and protocols in Quantum Information and Computation
\citep{Gardiner-Zoller,Scully-Zubairy}. Besides their origins in
Quantum Optics, they are also implemented in circuit QED, for example
\citep{cQED21}. Recently, the importance of third-order nonlinear
processes for completing a full set of quantum logic gates for quantum
computation in the context of continuous variables has been highlighted
\citep{GKP01,Zheng21,Hahn22}. However, mathematical difficulties
of the third-order squeezing operators or higher, such as divergent
matrix elements in the Fock basis states, were found many years ago
\citep{Fisher84}. Posterior developments avoided these issues, showing
that for evolutions in finite time, the states of these systems can
be calculated \citep{Braunstein87,Braunstein90,Ashhab25}. Furthermore,
it was demonstrated that finite and regular results are obtained for
models in which the squeezing terms are fed by quantized pumping operator,
that is, exchange Hamiltonians involving interactions between modes
with third-order or higher powers \citep{Hillery90,Drobny92,Tanas92,Banaszek97,Olsen02}.
Moreover, phase-space distribution representations were obtained,
which exhibit very suggestive symmetry structures, directly associated
with the order of the squeezing operators \citep{Braunstein87,Braunstein90,Tanas92,Banaszek97,Felbinger98}.
Within the context of the dissipative cavity systems, some works have
studied steady state properties, finding interesting bistability and
non-Gaussian features \citep{Felbinger98,Bajer91,Denys21}. 

Despite many experimental difficulties in handling third-order nonlinearities
in material media, three-photon nondegenerate down-conversion has
been achieved, by stimulating two output modes of a bulk crystal in
the traveling-wave regime \citep{Douady04,Bencheikh07,Gravier08}.
More recently, another version of the experiment has been performed
for a single stimulated output mode \citep{Bertrand25}. Meanwhile,
theoretical studies have modeled the operating conditions and shown
the quantum features of these processes \citep{Dot12,Borshchevskaya15,Okoth19,Bencheikh22}.
Another important achievement was the realization of spontaneous three-photon
parametric down-conversion in cavities of superconducting circuits
\citep{SandboChang20}. In that work, both degenerate and nondegenerate
cases were performed. It is worth mentioning active research in engineered
optical fibers, waveguides, and metamaterials for producing three-photon
parametric down-conversion processes in the traveling-wave regime
\citep{Corona11a,Corona11b,Akbari16,Cavanna16,Cavanna20,Banic22,Bacaoco25}. 

Considering the use for the diverse possibilities in quantum information
processing and computing, the three-photon down-conversion processes
can produce more than genuine tripartite entanglement, like the Greenberger–Horne–Zeilinger
(GHZ) states. In the case of the degenerate down-conversion, the nonlinear
unitary transformation depends on the third-order power of the creation
and annihilation operators, which has already been proposed as a way
to obtain a cubic-phase transformation, necessary for Gottesman-Kitaev-Preskill
(GKP) codes in the continuous-variable fault-tolerant quantum computation
\citep{GKP01,Zheng21,Hahn22}. Given these motivations, we have studied
the three-photon degenerate parametric down-conversion performed within
a cavity, taking into account certain realistic conditions, such as
the cavity dissipation and the stimulation of the generated output
photons by a driving coherent mode. Based on the experimental accomplishments
\citep{Bencheikh07,Bertrand25,SandboChang20}, we have considered
the pumping mode to be sufficiently intense, so that depletion effects
can be neglected and it can be treated as a classical wave. We have
focused on the stationary states of this system, obtaining finite
quadrature averages. However, unlike other driven dissipative systems
\citep{Felbinger98,Bajer91,Denys21}, we do not find multistable steady
states, but only a limited region of single stability, with transitions
to other unstable regions.

Furthermore, we have studied the spectra of quantum fluctuations of
the generated mode around the stable steady states. We have obtained
the spectral densities linked to correlation functions of the output
quadratures, revealing third-order squeezing and its frequency-dependent
structure, unlike the common squeezing. Due to the non-Gaussian nature
of the noise, we have also taken into account the third-order noise
components, which are commonly truncated, because of the representation
issues with the noise in terms of stochastic differential equations
\citep{Olsen02,Drummond14}. However, we have used an alternative
approach, namely the positive Wigner functions, developed by Drummond
\citep{Drummond14,Drummond17}. To calculate relevant quantities influenced
by these higher-order noises, we use three-time correlation functions
and calculate the respective two-dimensional frequency spectra, as
is already done in the magnetic resonance spectroscopy \citep{Ernst-Bodenhausen-Wokaun}.
In this way, we propose a theoretical tool to calculate the statistical
properties of non-Gaussian systems.

\section{Model}

Many variations of the three-photon down-conversion or triple photon
generation of nonlinear optical processes have been studied in the
literature. In this article, we focus on the original problem proposed
in theoretical works \citep{Fisher84,Braunstein87,Braunstein90,Ashhab25},
in which a generalization of squeezing by degenerate k-photon parametric
amplifiers was studied, but here we restrict ourselves to the three-photon
case. Moreover, we add the dynamics of the driven dissipative systems
to get closer to experimental achievements, where third-order down-conversion
implementations have been performed in recent years \citep{Douady04,Bencheikh07,Gravier08,Bertrand25,SandboChang20}.
In the physical model considered here, the nonlinear medium is fed
by a pumping laser mode within an optical cavity. The pumping mode
is sufficiently intense such that its depletion in down-conversion
process is negligible, allowing us to treat it as a classical pump
with frequency $\omega_{p}$. The triple photons produced by the parametric
down-conversion are degenerate with frequency $\omega_{0}$. A driving
mode with complex amplitude $F$ and frequency $\omega_{d}$ is added
for stimulating the triple photon mode. So the nonlinear process can
be modeled by the Hamiltonian 
\begin{equation}
H_{0}=\hbar\omega_{0}a^{\dagger}a+i\hbar(Fe^{-i\omega_{d}t}a^{\dagger}-F^{*}e^{i\omega_{d}t}a)+i\hbar\kappa(e^{-i\omega_{p}t}a^{\dagger3}-e^{i\omega_{p}t}a^{3}).\label{H0-semi-time}
\end{equation}
where $a$ is the annihilation operator of the subharmonic generated
mode a. The real constant parameter $\kappa$ characterizes the nonlinear
process and is proportional to the pumping mode amplitude.

To handle a time-independent Hamiltonian, we must perform a transformation
to the rotating frame. The frequencies of the pumping and driving
modes are externally controllable parameters, so we can set them as
$\omega_{p}=3\omega_{d}$. Thus the Hamiltonian (\ref{H0-semi-time})
can be reduced to 
\begin{equation}
H_{0}=\hbar\Delta a^{\dagger}a+i\hbar(Fa^{\dagger}-F^{*}a)+i\hbar\kappa(a^{\dagger3}-a^{3}),\label{H0-semi}
\end{equation}
 where we define the detuning $\Delta\equiv\omega_{0}-\omega_{d}=\omega_{0}-\omega_{p}/3$.

In addition to the cavity dynamics described by expression (\ref{H0-semi}),
the dissipative part is described by single-photon dissipator at zero
temperature,
\begin{equation}
\mathcal{D}(a)=\frac{\gamma}{2}\left(2a\rho a^{\dagger}-a^{\dagger}a\rho-\rho a^{\dagger}a\right).\label{dissipator}
\end{equation}
 Thus the density matrix of the cavity modes, $\rho$, is ruled by
the Lindbladian master equation \citep{Gardiner-Zoller}
\begin{equation}
\frac{d\rho}{dt}=-\frac{i}{\hbar}[H_{0},\rho]+\mathcal{D}(a).\label{master}
\end{equation}

In what follows, we analyze the stability of the system focusing on
the nonlinear interactions, so we must study the intracavity dynamics
described by $a$ and $a^{\dagger}$ operators and their combinations.
On the other hand, the study of spectral properties and noise levels
is more interesting for the modes that are actually detected, therefore
a method for calculating the extracavity modes is necessary. For this
task, we approached the issue using the input/output fields method
of Gardiner and Collett \citep{Gardiner-Zoller,Collett84,Gardiner85}.
In this way, we consider an input mode in the vacuum state, described
by annihilation operator $a^{\mathrm{(in)}}(t)$, the intracavity
modes with the respective operator $a(t)$, whose dynamics is ruled
by equation (\ref{master}), and output mode, with respective operator
$a^{\mathrm{(out)}}(t)$, which follows the condition
\begin{equation}
a^{\mathrm{(out)}}(t)=a^{\mathrm{(in)}}(t)+\sqrt{\gamma}a(t).\label{in-out}
\end{equation}
To focus in the nonlinear process, we model the system as a one-sided
cavity, that is, the cavity has relevant loss through only one mirror,
whereas the pumping and driving modes are injected through other cavity
ports without substantial losses. 

In the spectral study of the system, we apply the Fourier analysis
to the respective stochastic variables of the intracavity mode and
we model the input mode with $[a^{\mathrm{(in)}}(t),a^{\mathrm{(in)\dagger}}(t^{\prime})]=\delta(t-t^{\prime})$,
so that in next Section we associate it with a Wiener process (or,
for the sake of integration, a white noise). Moreover, the input mode
is understood as the resultant bath modes incoming to cavity, so we
assume it as being in the vacuum state. In this way, we can obtain
the spectral densities for the output mode by Eq. (\ref{in-out}). 

\section{Equations in phase-space}

By standard methods \citep{Gardiner}, the master equation (\ref{master})
can be mapped into a partial differential equation for the Wigner
function $W$, known as Kramers-Moyal equation (KME), taking the form
\begin{eqnarray}
\frac{\partial W(\alpha,\alpha^{*})}{\partial t} & = & \left\{ \frac{\partial}{\partial\alpha}\left[\Big(\frac{\gamma}{2}+i\Delta\Big)\alpha-F-3\kappa\alpha^{*2}\right]+\frac{\partial}{\partial\alpha^{*}}\left[\Big(\frac{\gamma}{2}-i\Delta\Big)\alpha^{*}-F^{*}-3\kappa\alpha^{2}\right]\right.\label{wig}\\
 & + & \left.\frac{\gamma}{2}\frac{\partial^{2}}{\partial\alpha\partial\alpha^{*}}-\frac{\kappa}{4}\left(\frac{\partial^{3}}{\partial\alpha^{3}}+\frac{\partial^{3}}{\partial\alpha^{*3}}\right)\right\} W(\alpha,\alpha^{*})
\end{eqnarray}
where the ordinary complex variables $\alpha$ and $\alpha^{*}$ are
analogous to the operators $a$ and $a^{\dagger}$ in the domain of
the Wigner function. Third or higher-order equations, as equation
(\ref{wig}), do not have a direct correspondence with stochastic
differential equations, because, in general, their solutions are not
always positive to any conditions, so they are not genuine probability
densities \citep{Olsen02,Drummond14,Pawula67}. A method developed
by Drummond to avoid this issue is to extend the domain of the $W$
function, so that the complex variables $\alpha$ and $\alpha^{*}$
become independent complex variables, not complex conjugate to each
other, as $\alpha\rightarrow\alpha$ and $\alpha^{*}\rightarrow\beta$.
Thus, the dimension of the original domain doubles to become a two-dimensional
complex domain, bringing the new $W$ function to be necessarily positive,
with analytic moments equivalent to those of the original Wigner function
\citep{Drummond14,Drummond17}. This procedure follows closely the
well-known positive-P function formalism \citep{Drummond80a}, for
this reason the new $W$ is called positive-W function. 

The correspondence between the Fokker-Planck equations and the stochastic
differential equations, obtained by the relations of the Itô's calculus,
is well-known \citep{Gardiner}. Following Drummond \citep{Drummond17},
we associate the KME (\ref{wig}) to a stochastic differential equation,
extending the domain to a double complex space and, in particular,
transforming the third-order derivatives in Eq. (\ref{wig}) into
third-order noises. Thus, the differential equations for respective
stochastic variables $\alpha(t)$ and $\beta(t)$ arranged in vector
form, with their variations $dx=(d\alpha(t),d\beta(t))^{T}$, are
\begin{equation}
dx=Ddt+R^{(2)}dZ+R^{(3)}d\Xi,\label{eq-a}
\end{equation}
 such that the second and third-order noises are $dZ=(dZ_{1},dZ_{2})^{T}$
and $d\Xi=(d\Xi_{1},d\Xi_{2})^{T}$, respectively. The drift vector
$D$ is
\begin{equation}
D=\left(\begin{array}{c}
(-\frac{\gamma}{2}-i\Delta)\alpha+F+3\kappa\beta^{2}\\
(-\frac{\gamma}{2}+i\Delta)\beta+F^{*}+3\kappa\alpha^{2}
\end{array}\right)\label{drift-a}
\end{equation}
 and the noise coefficients are matrices $R^{(2)}=\sqrt{\frac{\gamma}{4}}\left(\begin{array}{cc}
1 & i\\
1 & -i
\end{array}\right)$ and $R^{(3)}=\sqrt[3]{\frac{3\kappa}{8}}\left(\begin{array}{cc}
1 & 0\\
0 & 1
\end{array}\right)$. The noises are characterized by their correlations: 
\begin{equation}
\left\langle dZ_{i}\right\rangle =0,\label{2noise-mean-a}
\end{equation}
\begin{equation}
\left\langle dZ_{i}dZ_{j}\right\rangle =\delta_{ij}dt,\label{2noise-corr-a}
\end{equation}
to the second-order noises and 
\begin{equation}
\left\langle d\Xi_{i}\right\rangle =\left\langle d\Xi_{i}d\Xi_{j}\right\rangle =0,\label{3noise-mean-a}
\end{equation}
\begin{equation}
\left\langle d\Xi_{i}d\Xi_{j}d\Xi_{k}\right\rangle =\delta_{ij}\delta_{jk}dt,\label{3noise-corr-a}
\end{equation}
to the third-order ones. The system proposed in this article brings
an advantage in use of the Drummond's method because, whereas the
KME second-order derivatives are non-diagonal but can be treated in
the usual way as in the Fokker-Planck equation, the third-order derivatives
correspond to the diagonal noise case, whose correlations follow a
simple and direct form as in Eqs. (\ref{2noise-mean-a}) and (\ref{2noise-corr-a}).
We can also note that the third-order noises are evidently non-Gaussian
contributions to the stochastic behavior of the system and they have
origin only from the nonlinear three-photon generation.

To study the squeezing and the respective spectrum of the three-photon
down-conversion, it is useful to perform a change of variables to
quadrature phase amplitudes, namely, 
\begin{eqnarray}
q & = & \frac{1}{2}(\alpha+\beta),\label{quadratures}\\
p & = & \frac{1}{2i}(\alpha-\beta),\nonumber 
\end{eqnarray}
 so that the respective stochastic quadrature variables obey the following
equations 
\begin{equation}
dq=\left(-\frac{\gamma}{2}q+\Delta p+\Re(F)+3\kappa(q^{2}-p^{2})\right)dt+\frac{\sqrt{\gamma}}{2}Z_{1}+\frac{\sqrt[3]{3\kappa}}{2}\Xi_{+}\label{eq-q}
\end{equation}
 and
\begin{equation}
dp=\left(-\frac{\gamma}{2}p-\Delta q+\Im(F)-6\kappa qp\right)dt+\frac{\sqrt{\gamma}}{2}Z_{2}-i\frac{\sqrt[3]{3\kappa}}{2}\Xi_{-}.\label{eq-p}
\end{equation}
 Note that the second-order noise term is diagonal for the new variables
of quadratures, whereas the third-order one is not. Even so, we can
redefine the third-order noises as
\begin{equation}
\Xi_{+}=\frac{\Xi_{1}+\Xi_{2}}{\sqrt[3]{2}}\label{3noise-plus}
\end{equation}
and
\begin{equation}
\Xi_{-}=\frac{\Xi_{1}-\Xi_{2}}{\sqrt[3]{2}},\label{3noise-minus}
\end{equation}
 whose correlation functions can be found as
\begin{equation}
\left\langle d\Xi_{+}d\Xi_{+}d\Xi_{+}\right\rangle =dt\label{3noise-corr-plus}
\end{equation}
and
\begin{equation}
\left\langle d\Xi_{+}d\Xi_{-}d\Xi_{-}\right\rangle =dt\label{3noise-corr-pm}
\end{equation}
to the nonvanishing cases. The other null correlations are 
\begin{equation}
\left\langle d\Xi_{-}d\Xi_{-}d\Xi_{-}\right\rangle =\left\langle d\Xi_{+}d\Xi_{+}d\Xi_{-}\right\rangle =0\label{3noise-corr-null}
\end{equation}
Looking at Eqs. (\ref{3noise-corr-plus})–(\ref{3noise-corr-null}),
we note the non-trivial combination of noises for null and non-null
correlation functions. This is consequence of the non-diagonal third-order
noises for the differential quadrature equations (\ref{eq-q}) and
(\ref{eq-p}) \citep{Drummond17}.

\section{Steady state conditions}

Deterministic semi-classical equations of the quadrature variables
can be obtained by calculating the expectations of the stochastic
equations and applying the mean-field approximation, that is, in the
present calculation, averages of variable products are factored as
products of averages. In the stationary state, the stochastic variations
vanish in the mean, so that the equations reduce to 
\begin{equation}
-\frac{\gamma}{2}\left\langle q\right\rangle _{ss}+\Delta\left\langle p\right\rangle _{ss}+\Re(F)+3\kappa\left(\left\langle q\right\rangle _{ss}^{2}-\left\langle p\right\rangle _{ss}^{2}\right)=0\label{ss-q}
\end{equation}
 and
\begin{equation}
-\frac{\gamma}{2}\left\langle p\right\rangle _{ss}-\Delta\left\langle q\right\rangle _{ss}+\Im(F)-6\kappa\left\langle q\right\rangle _{ss}\left\langle p\right\rangle _{ss}=0,\label{ss-p}
\end{equation}
where the $ss$ subscript denotes the steady state solutions.

\begin{figure*}
\begin{centering}
\includegraphics[width=0.3\textwidth]{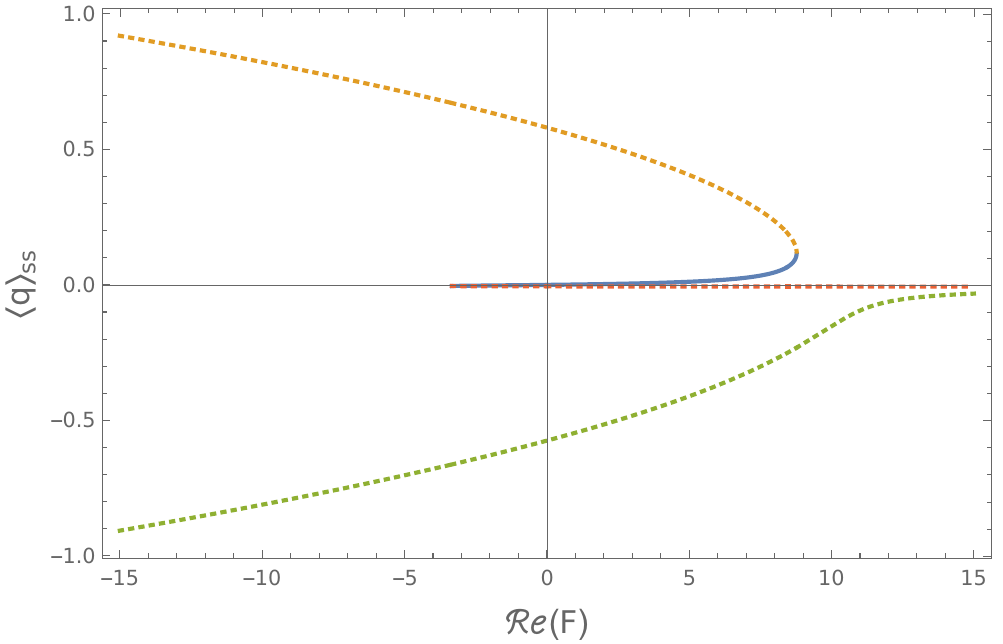}\quad{}\includegraphics[width=0.3\textwidth]{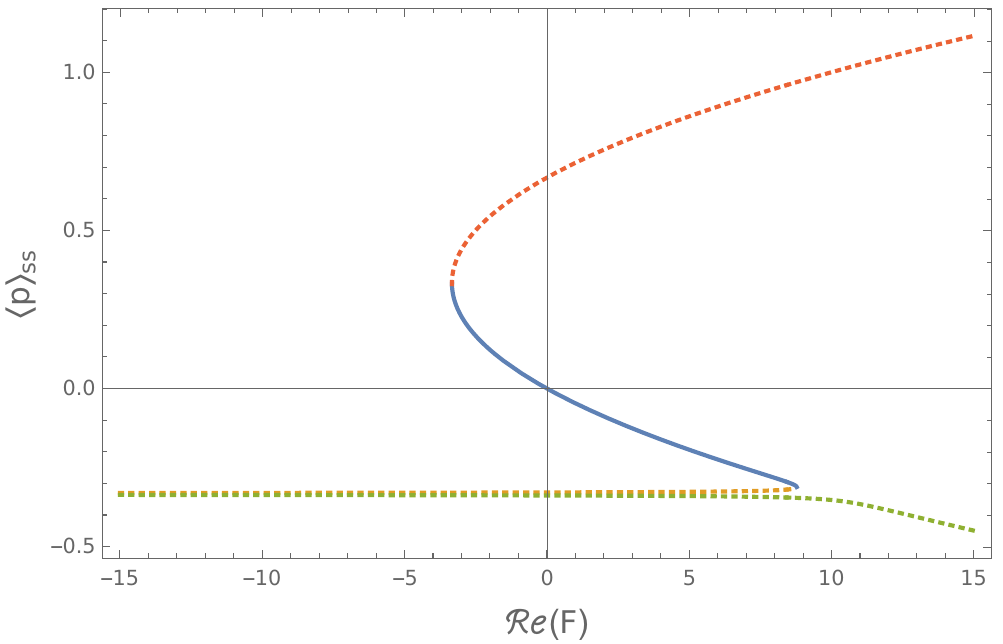}\quad{}\includegraphics[width=0.3\textwidth]{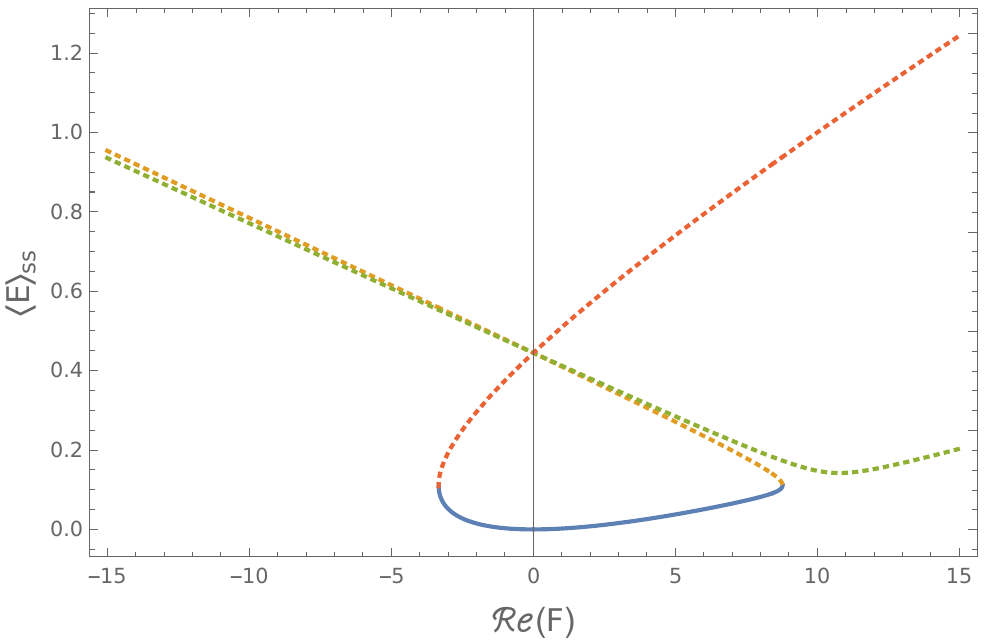}
\par\end{centering}
\caption{Steady state average values of the quadratures $\langle q\rangle_{ss}$
and $\langle p\rangle_{ss}$, and of the respective semi-classical
energy $\langle E\rangle_{ss}=\langle q\rangle_{ss}^{2}+\langle p\rangle_{ss}^{2}$,
as functions of the real part of the driving mode amplitude, $\Re(F)$,
for the four physical solutions. Blue continuous lines represent the
unique stable steady branch. The three other dashed lines represent
unstable steady branches. The parameters for the three plots are $\gamma=1$,
$\Delta=20$, $\kappa=10$, and $\Im(F)=0$. \protect\label{bifurcation-real-F}}
\end{figure*}

\begin{figure*}
\begin{centering}
\includegraphics[width=0.3\textwidth]{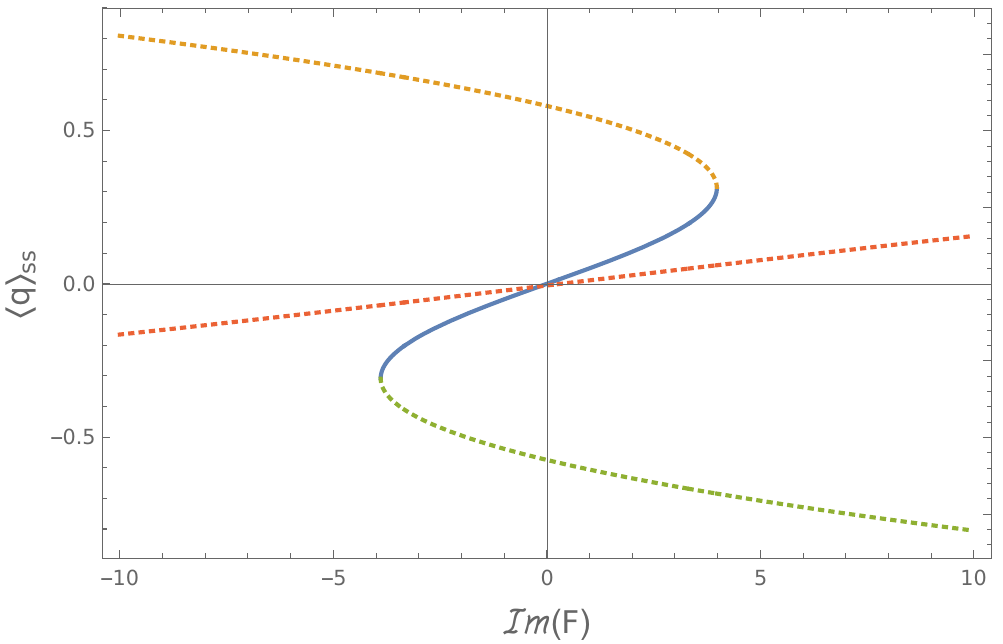}\quad{}\includegraphics[width=0.3\textwidth]{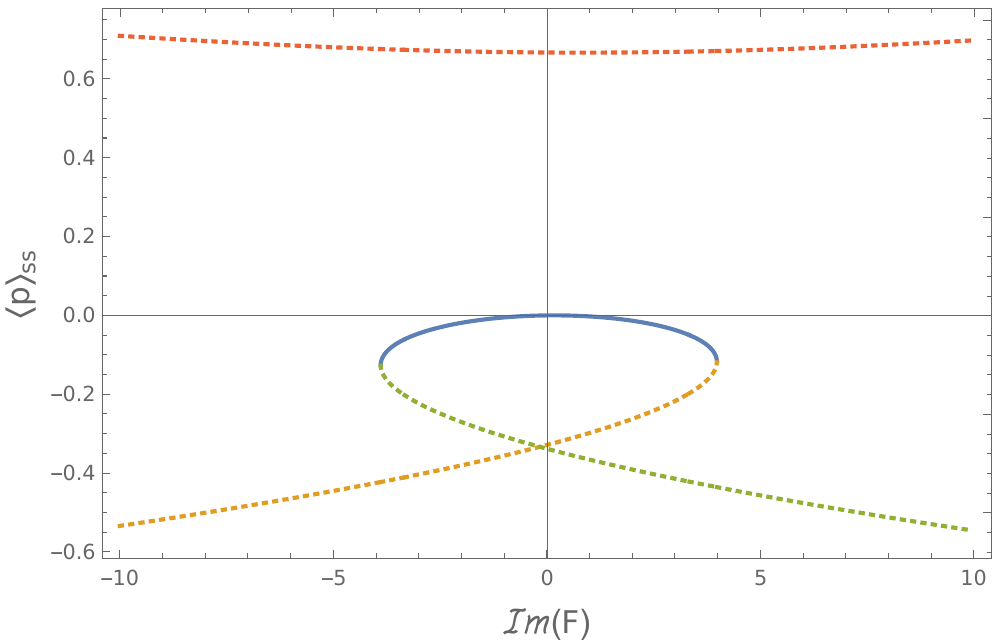}\quad{}\includegraphics[width=0.3\textwidth]{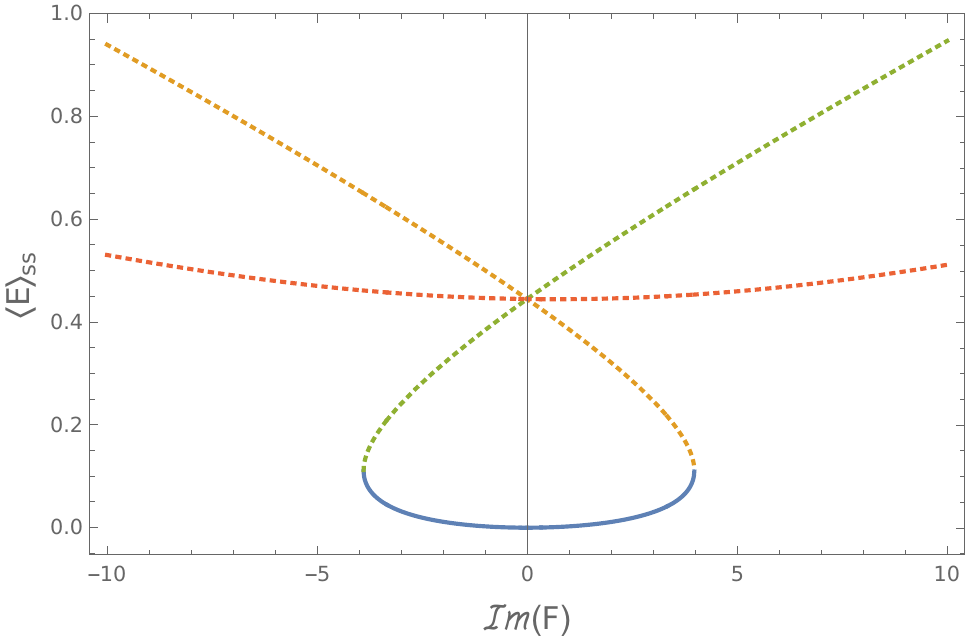}
\par\end{centering}
\caption{Steady state average values of the quadratures $\langle q\rangle_{ss}$
and $\langle p\rangle_{ss}$, and of the respective semi-classical
energy $\langle E\rangle_{ss}=\langle q\rangle_{ss}^{2}+\langle p\rangle_{ss}^{2}$,
as functions of the imaginary part of the driving mode amplitude,
$\Im(F)$, for the four physical solutions. Blue continuous lines
represent the unique stable steady branch. The three other dashed
lines represent unstable steady branches. The parameters for the three
plots are $\gamma=1$, $\Delta=20$, $\kappa=10$, and $\Re(F)=0$.\protect\label{bifurcation-imaginary-F}}
\end{figure*}

We can decouple Eqs. (\ref{ss-q}) and (\ref{ss-p}), to obtain two
independent quartic algebraic equations for both mean quadratures.
Their solutions are very cumbersome and will be omitted. With these
results, we can plot bifurcation diagrams and visualize the stable
and unstable steady state branches. In Fig. \ref{bifurcation-real-F},
we show the bifurcation diagrams of the mean quadratures $\left\langle q\right\rangle _{ss}$,
$\left\langle p\right\rangle _{ss}$, and the semi-classical energy
$\left\langle E\right\rangle _{ss}=\left\langle q\right\rangle _{ss}^{2}+\left\langle p\right\rangle _{ss}^{2}$
as functions of $\Re(F)$, for the case with $\Im(F)=0$. In Fig.
\ref{bifurcation-imaginary-F}, we show the same diagrams, but the
quantities are as functions of $\Im(F)$ and setting $\Re(F)=0$.

In both Figs. \ref{bifurcation-real-F} and \ref{bifurcation-imaginary-F},
the stable steady states are represented only by blue lines, all other
dotted lines represent unstable steady states. It is practically impossible
for these unstable states to be attainable, so that the possibly observable
steady states are restricted to a bounded set of parameters. Another
aspect notable is the difference and the asymmetry of the diagrams,
depending on the real and imaginary values of the driving mode. Such
system's features can be better understood analyzing its stability
conditions \citep{Strogatz}. 

The stability of the branches presented in Figs. \ref{bifurcation-real-F}
and \ref{bifurcation-imaginary-F} can be established by the Jacobian
of the average of the stochastic equations (\ref{eq-q}) and (\ref{eq-p}),
\begin{equation}
J=\begin{pmatrix}\partial\langle\dot{q}\rangle/\partial\langle q\rangle & \partial\langle\dot{q}\rangle/\partial\langle p\rangle\\
\partial\langle\dot{p}\rangle/\partial\langle q\rangle & \partial\langle\dot{p}\rangle/\partial\langle p\rangle
\end{pmatrix},\label{jac}
\end{equation}
hence
\begin{equation}
J=\begin{pmatrix}-\frac{\gamma}{2}+6\kappa\langle q\rangle & \Delta-6\kappa\langle p\rangle\\
-\Delta-6\kappa\langle p\rangle & -\frac{\gamma}{2}-6\kappa\langle q\rangle
\end{pmatrix}.\label{jac-calc}
\end{equation}

For the identification of the stable or unstable steady states, the
Jacobian must obey the following inequalities 
\begin{equation}
\mathrm{tr}J=-\gamma\leq0\label{trj}
\end{equation}
and 
\begin{equation}
\det J=\left(\frac{\gamma}{2}\right)^{2}+\Delta^{2}-36\kappa^{2}\left(\langle q\rangle^{2}+\langle p\rangle^{2}\right)\geq0,\label{detj}
\end{equation}
calculated to the solutions found from Eqs. (\ref{ss-q}) and (\ref{ss-p}).
The first condition is clearly always met, but the second condition
comprises a limited region. In this way, the stable and unstable steady
state branches can be established in Figs. \ref{bifurcation-real-F}
and \ref{bifurcation-imaginary-F}.

\begin{figure*}
\begin{centering}
\includegraphics[width=0.3\textwidth]{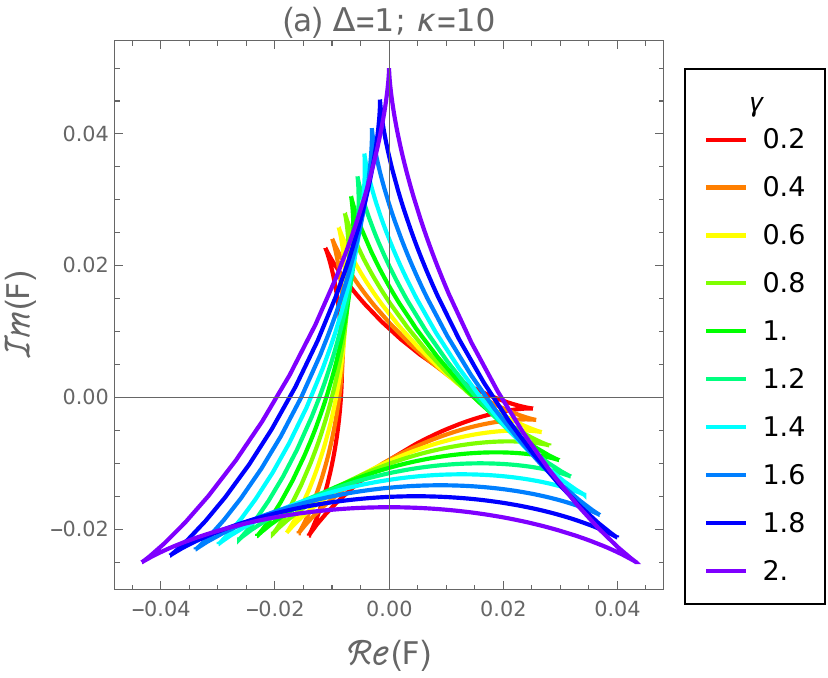}\quad{}\includegraphics[width=0.3\textwidth]{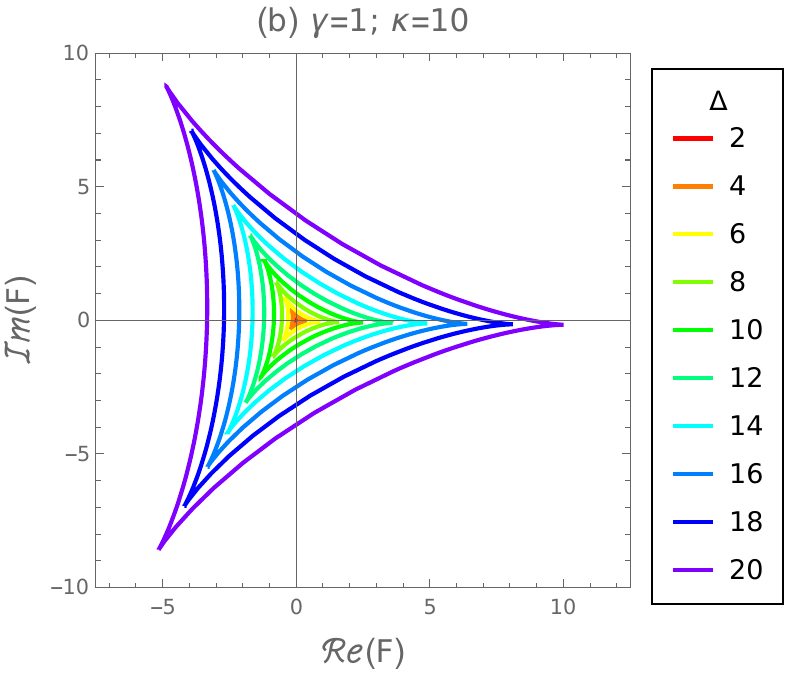}\quad{}\includegraphics[width=0.3\textwidth]{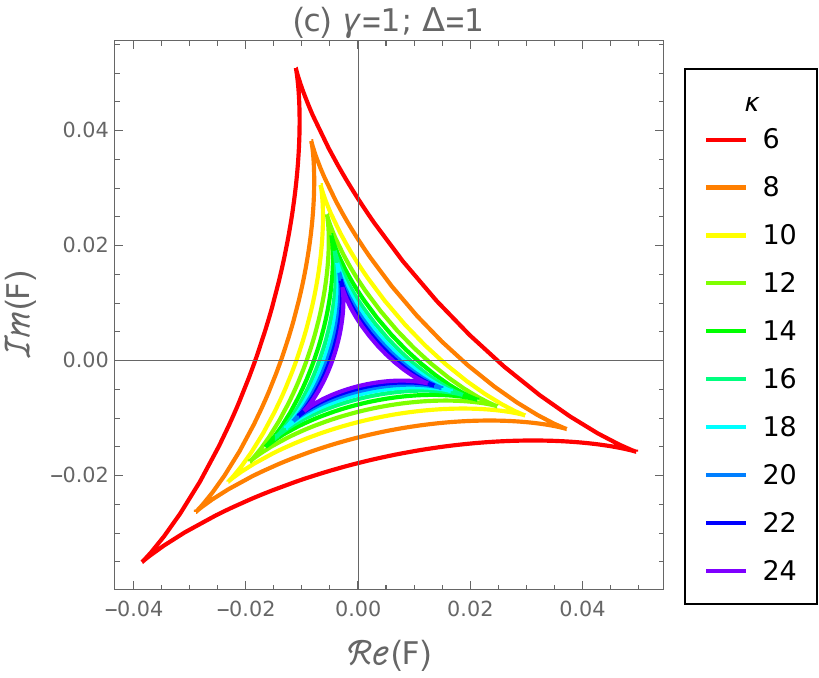}
\par\end{centering}
\caption{Stability region boundary as function of the real and imaginary parts
of the driving mode amplitude. The driving mode amplitude values for
which the system exhibits stable steady states lie within the region
bounded by the contours. In plot (a), we fixed the detuning $\Delta=1$
and the nonlinear coupling constant $\kappa=10$ and varied the dissipative
factor $\gamma$. In plot (b), we fixed $\gamma=1$ and $\kappa=10$
and varied $\Delta$. In plot (c), we fixed $\gamma=1$ and $\Delta=1$
and varied $\kappa$. \protect\label{regions}}
\end{figure*}

The boundary of the stability condition is found with $\det(J_{b})=0.$
Rewriting the quadrature variables in terms of polar variables, $q_{b}=r_{b}\cos\theta$
and $p_{b}=r_{b}\sin\theta$, the stability boundary is found as 
\begin{equation}
q_{b}^{2}+p_{b}^{2}=r_{b}^{2}=\frac{\left(\frac{\gamma}{2}\right)^{2}+\Delta^{2}}{(6\kappa)^{2}}.\label{esta}
\end{equation}
 Substituting in the Eqs. (\ref{ss-q}) and (\ref{ss-p}), we obtain
the stability boundary for the real and imaginary parts of the driving
mode amplitude, 
\begin{equation}
\Re(F_{b})=r_{b}\left[\frac{\gamma}{2}\cos\theta-\Delta\sin\theta-3\kappa r_{b}\left(\cos^{2}\theta-\sin^{2}\theta\right)\right]\label{re}
\end{equation}
 and 
\begin{equation}
\Im(F_{b})=r_{b}\left(\frac{\gamma}{2}\sin\theta+\Delta\cos\theta+6\kappa r_{b}\cos\theta\sin\theta\right)\label{im}
\end{equation}
 Expressions (\ref{re}) and (\ref{im}) can be plotted in a complex
plane of the driving mode amplitude, so that the inner region of the
boundary line represents the stability region, as we see in Fig. \ref{regions},
varying the parameters $\gamma$, $\Delta$ and $\kappa$.

The triply folded aspect of the stability region reveals an asymmetric
projection onto both components of the driving mode amplitude, $\Re(F)$
and $\Im(F)$, reflecting the asymmetries in Figs. \ref{bifurcation-real-F}
and \ref{bifurcation-imaginary-F}. 

To complement our stability analysis, let us investigate the existence
of steady states regarding the photon number of the mode generated
within the cavity. We follow previous analyses which utilized the
Heisenberg equation for the isolated system \citep{Hillery90,Olsen02}.
However, given the inclusion of the quantum mode dissipation, we must
account for a modification to the Heisenberg equation. To this goal,
we consider the input-output field method \citep{Gardiner-Zoller,Collett84,Gardiner85}.
Thus, any operator $\mathcal{O}$ associated with the generated mode
must satisfy the equation
\begin{equation}
\frac{d\mathcal{O}}{dt}=-\frac{1}{\hbar}[\mathcal{O},H_{0}]-\left[[\mathcal{O},a^{\dagger}]\left(\frac{\gamma}{2}a+\sqrt{\gamma}a^{\mathrm{(in)}}\right)-\left(\frac{\gamma}{2}a^{\dagger}+\sqrt{\gamma}a^{\mathrm{(in)}\dagger}\right)[\mathcal{O},a]\right].\label{eq-in-out}
\end{equation}
To analyze the photon number evolution, we consider the operator $\mathcal{O}\rightarrow a^{\dagger}a$
and calculate up to its second time derivative. We obtain terms that
depend purely on the intracavity mode operators, or products of them
with the input modes, or a pure input term involving $a^{\mathrm{(in)}\dagger}a^{\mathrm{(in)}}$.
Assuming the input mode bath is in the vacuum state, and the intracavity
operators are reasonably described as non-anticipating quantities,
so that they are uncorrelated to the input operators, then all the
terms with the input mode operators must cancel out. 

\begin{figure}
\begin{centering}
\includegraphics[width=0.5\columnwidth]{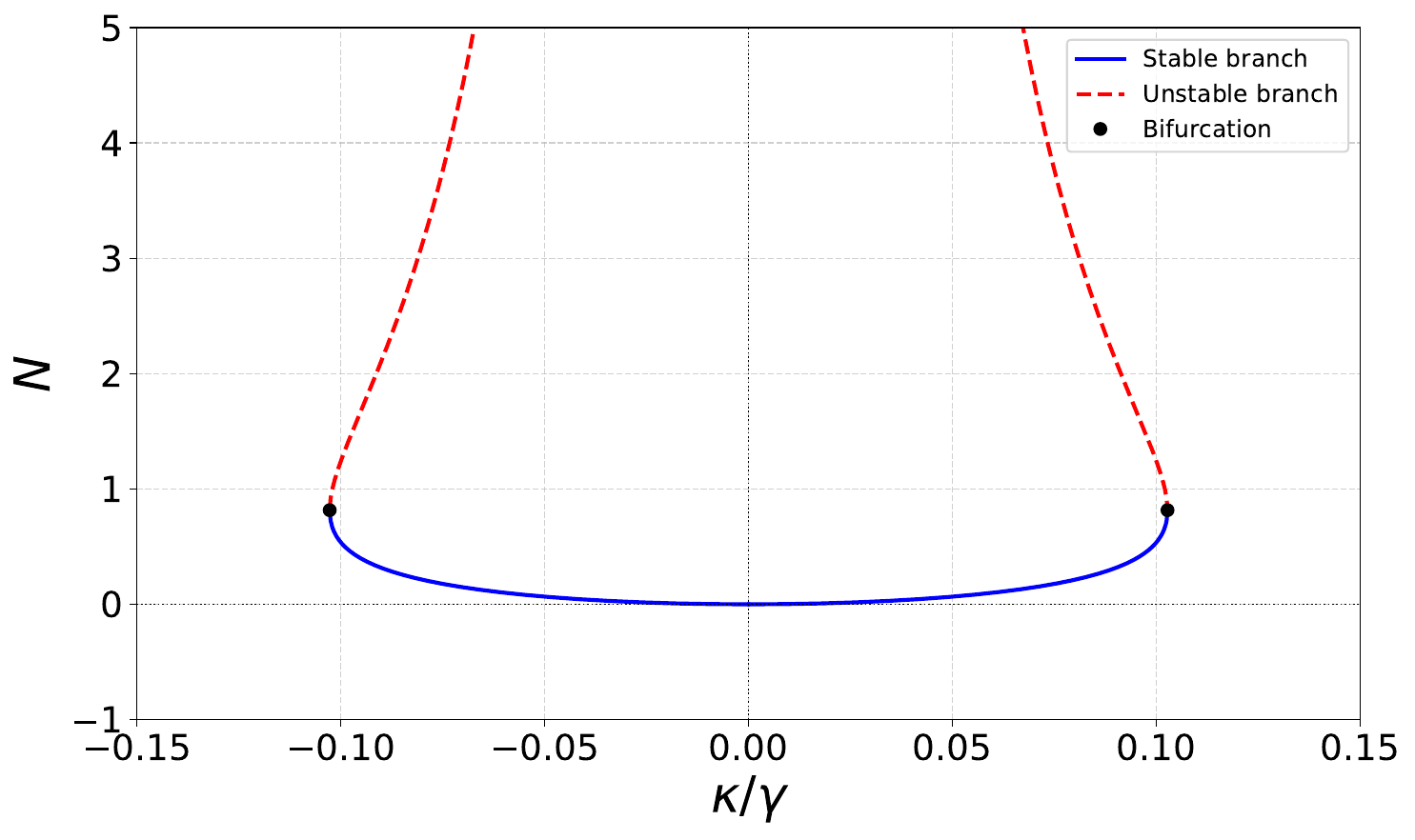}
\par\end{centering}
\caption{Mean photon number to steady states without driving mode, $F=0$.
Blue continuous line represents the unique stable steady state branch.
Red dashed lines represent the unstable steady states. \protect\label{bifurcation-number}}
\end{figure}

Again, making the mean-field approximation, but here assuming that
all operator products above the second power factorize, the equation
for mean values is 
\begin{eqnarray}
\frac{d^{2}\mathcal{N}}{dt} & = & i\Delta\left[\mathcal{N},\frac{d\mathcal{N}}{dt}\right]-\frac{5\gamma}{2}\frac{d\mathcal{N}}{dt}+2|F|^{2}+\gamma\left(F^{*}\langle a\rangle+F\langle a^{\dagger}\rangle\right)+12\kappa\left(F\langle a^{2}\rangle+F^{*}\langle a^{\dagger2}\rangle\right)+\nonumber \\
 & + & 18\kappa^{2}(3\mathcal{N}^{2}+3\mathcal{N}+2)-\frac{3\gamma^{2}}{2}\mathcal{N}.\label{mean-number-eq}
\end{eqnarray}
where we define $\mathcal{N}=\langle a^{\dagger}a\rangle$. Of course
this equation is incomplete, because it involves independent terms
as $\langle a\rangle$ and $\langle a^{2}\rangle$ and any attempt
to add other equations would result an infinite equation system. However,
for the sake of comparison let us discard the driving mode, that is
$F=0$, since it was not considered in the previous articles. In the
steady state, all time derivatives vanish, yielding the algebraic
equation for the respective mean photon number 
\begin{equation}
18\kappa^{2}(3\mathcal{N}_{ss}^{2}+3\mathcal{N}_{ss}+2)-\frac{3\gamma^{2}}{2}\mathcal{N}_{ss}=0.\label{mean-number-ss}
\end{equation}

We have plotted a bifurcation diagram of Eq. (\ref{mean-number-ss}),
as it is shown in Fig. \ref{bifurcation-number}. We find that, unlike
in previous studies \citep{Hillery90,Olsen02}, the average photon
number converges asymptotically to a branch of stable steady states,
as shown by the solid blue line in Fig. \ref{bifurcation-number}.
The difference here lies in the contribution of the dissipative term
in Eq. (\ref{mean-number-ss}), which regularizes the system behavior
and keeps the photon number finite. Interestingly, it can be observed
that outside the stable steady state region, the average photon number
diverges in finite time, revealing a remnant of the highly irregular
behavior associated with three-photon down-conversion, a widely studied
aspect \citep{Fisher84,Braunstein87,Hillery90,Olsen02}.

\section{Spectra in the stable states}

An analysis of the spectral densities of the generated three-photon
mode can be done for steady states using standard Fourier transform
methods. The dynamical properties of the system in the neighborhood
of its fixed points are sufficiently evaluated by linearizing around
those points, thus it is pertinent to apply the Fourier analysis for
finding stationary spectra near the stable fixed points \citep{Gardiner,Strogatz}. 

Defining a vector of the quadrature variables centered in the stable
steady states (sss),
\begin{equation}
Q=\begin{pmatrix}Q_{q}\\
Q_{p}
\end{pmatrix}=\begin{pmatrix}q-\langle q\rangle_{sss}\\
p-\langle p\rangle_{sss}
\end{pmatrix},\label{quad-center}
\end{equation}
 we can obtain the linearized stochastic differential equation calculating
the Jacobian (\ref{jac-calc}) to the stable steady branch, $A=J|_{sss}$.
In the following analysis, it is suitable to use the Langevin equation
formalism, since it is more natural to the calculations of the Fourier
analysis. Hence, we get
\begin{equation}
\frac{dQ}{dt}=AQ+B\zeta+C\xi,\label{langevin}
\end{equation}
where $B$ and $C$ are matrix coefficients of the noise terms, given
by
\begin{equation}
B=\sqrt{\frac{\gamma}{4}}\begin{pmatrix}1 & 0\\
0 & 1
\end{pmatrix}\label{coefficient-b}
\end{equation}
 and 
\begin{equation}
C=\sqrt[3]{\frac{3\kappa}{8}}\begin{pmatrix}1 & 0\\
0 & -i
\end{pmatrix}.\label{coefficient-c}
\end{equation}
 The noises in the Langevin equation (\ref{langevin}) are obtained
by matching the mean value calculations of stochastic variables, or
using Ito´s integration with Wiener process or using the ordinary
calculus with white noise, that is, we follow the prescription done
by Gardiner in reference \citep{Gardiner}. Roughly speaking, we take
$dZ_{i}\rightarrow\zeta_{i}dt$ and $d\Xi_{i}\rightarrow\xi_{i}dt$.
In this way, the noise vectors are 
\begin{equation}
\zeta=\begin{pmatrix}\zeta_{1}\\
\zeta_{2}
\end{pmatrix}\quad,\quad\xi=\begin{pmatrix}\xi_{+}\\
\xi_{-}
\end{pmatrix},\label{noises}
\end{equation}
 characterized by the following mean values, $\langle\zeta_{i}(t)\rangle=\langle\xi_{i}(t)\rangle=0$.
The second-order noises obey the correlations 
\begin{equation}
\langle\zeta_{i}(t)\zeta_{j}(t^{\prime})\rangle=\delta_{ij}\delta(t-t^{\prime}).\label{2noise-langevin}
\end{equation}
 To the third-order noise, we must obtain (see Section 7.7 in \citep{Gardiner}
and \citep{Gardiner77})
\begin{equation}
\langle\xi_{+}(t)\xi_{+}(t^{\prime})\xi_{+}(t^{\prime\prime})\rangle=\langle\xi_{+}(t)\xi_{-}(t^{\prime})\xi_{-}(t^{\prime\prime})\rangle=\delta(t-t^{\prime})\delta(t-t^{\prime\prime})\label{3noise-langevin-delta}
\end{equation}
 and 
\begin{equation}
\langle\xi_{+}(t)\xi_{+}(t^{\prime})\xi_{-}(t^{\prime\prime})\rangle=\langle\xi_{-}(t)\xi_{-}(t^{\prime})\xi_{-}(t^{\prime\prime})\rangle=0.\label{3noise-langevin-null}
\end{equation}

The Fourier transforms of the centered quadrature variables are
\begin{equation}
\tilde{Q}_{i}(\omega)=\frac{1}{2\pi}\int_{-\infty}^{+\infty}Q_{i}(t)e^{-i\omega t}dt,\label{quad-fourier}
\end{equation}
 and the respective inverse transforms are 
\begin{equation}
Q_{i}(t)=\int_{-\infty}^{+\infty}\tilde{Q_{i}}(\omega)e^{i\omega t}d\omega.\label{quad-inverse}
\end{equation}
Moreover, the relevant spectral densities are not concerning the intracavity
variables, but are those related with the output mode. Based on Eq.
(\ref{in-out}), we have in terms of quadrature variables
\begin{equation}
Q^{\mathrm{(out)}}(t)=Q^{\mathrm{(in)}}(t)+\sqrt{\gamma}Q(t),\label{quad-in-out}
\end{equation}
 where the input variable vector is $Q^{\mathrm{(in)}}(t)=(q^{\mathrm{(in)}}(t),p^{\mathrm{(in)}}(t))^{T}=(\zeta_{1}(t),\zeta_{2}(t))^{T}$,
that is, the input mode is just the second-order noise injected into
the cavity. Thus the Fourier transform of Eq. (\ref{quad-in-out})
is 
\begin{equation}
\tilde{Q}^{\mathrm{(out)}}(\omega)=\tilde{Q}^{\mathrm{(in)}}(\omega)+\sqrt{\gamma}\tilde{Q}(\omega),\label{in-out-fourier}
\end{equation}
 so that the input mode is considered to be in the vacuum state, as
presented in Section II. Hence, $\tilde{Q}_{i}^{\mathrm{(in)}}(\omega)=\tilde{\zeta}_{i}(\omega)$
with $\left\langle \tilde{\zeta}_{i}(\omega)\tilde{\zeta}_{j}^{*}(\omega^{\prime})\right\rangle =\delta_{ij}\delta(\omega-\omega^{\prime})$
\citep{Gardiner-Zoller,Gardiner85}. Thus the two-time correlation
functions related to the output mode are 
\begin{equation}
\phi_{ij}(\tau)=\left\langle Q_{i}^{\mathrm{(out)}}(t)Q_{j}^{\mathrm{(out)}}(t+\tau)\right\rangle ,\label{2correlation-quad}
\end{equation}
 and their Fourier transforms define the respective spectral densities
\citep{Gardiner},
\begin{equation}
S_{ij}(\omega)=\frac{1}{2\pi}\int_{-\infty}^{+\infty}\phi_{ij}(\tau)e^{-i\omega\tau}d\tau.\label{2spectrum}
\end{equation}
 The spectral densities are also related to the correlation functions
of the transformed quantities
\begin{equation}
\left\langle \tilde{Q_{i}}^{\mathrm{(out)}}(\omega)\tilde{Q_{j}}^{\mathrm{(out)}*}(\omega^{\prime})\right\rangle =S_{ij}(\omega)\delta(\omega-\omega^{\prime}),\label{2correlation-fourier}
\end{equation}
of which we note that $S_{ij}(\omega)=S_{ji}(\omega)^{*}$. For notation
simplicity, we use $S_{ii}(\omega)\equiv S_{i}(\omega)$.

\begin{figure}
\begin{centering}
\includegraphics[width=0.5\columnwidth]{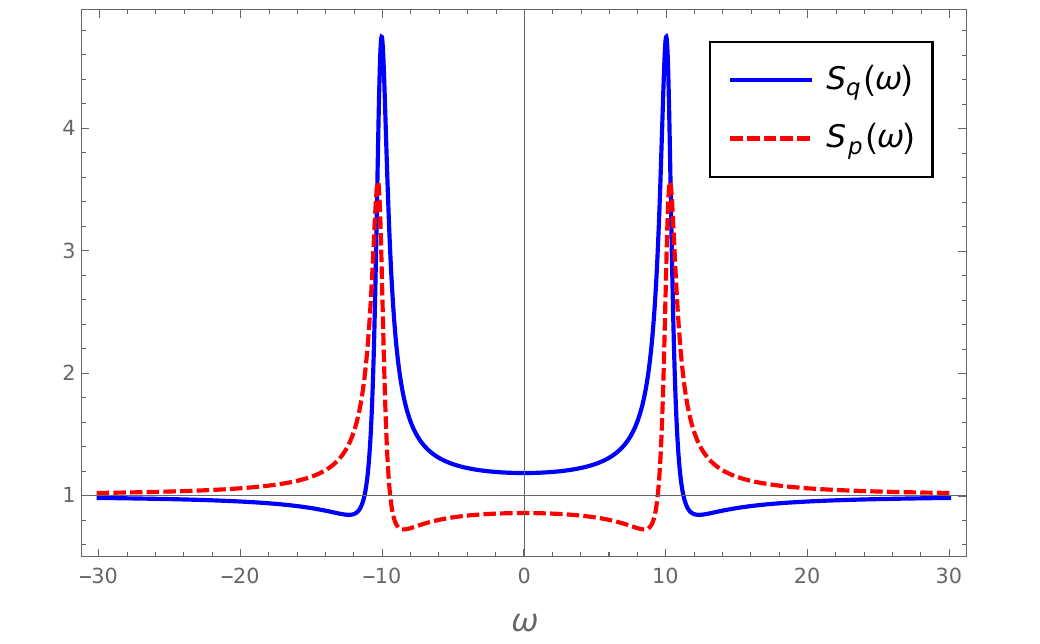}
\par\end{centering}
\caption{Noise spectra of the quadratures $Q_{q}^{\mathrm{(out)}}$ and $Q_{p}^{\mathrm{(out)}}$
of the output modes from the cavity. Respectively, the blue continuous
curve is the spectrum $S_{q}(\omega)$ and the red dashed curve is
the spectrum $S_{p}(\omega)$. The parameters for the plots are $\gamma=1$,
$\Delta=20$, $\kappa=10$, and $F=i3.9$. SQL is set to 1. \protect\label{squeezing-spectrum}}
\end{figure}

\begin{figure}
\begin{centering}
\includegraphics[width=0.5\columnwidth]{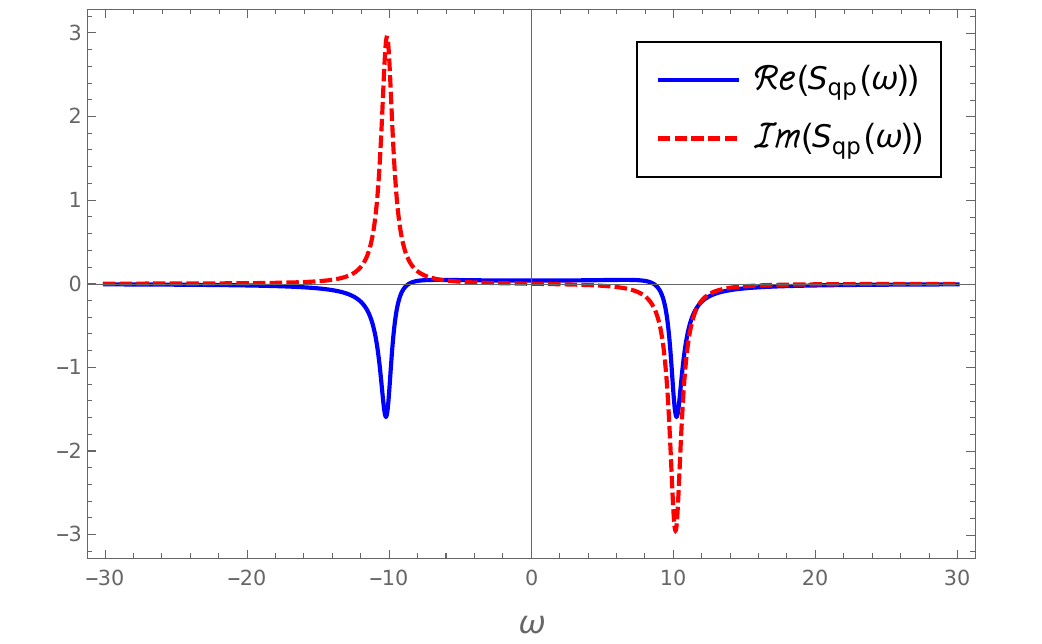}
\par\end{centering}
\caption{Cross noise spectrum of the quadrature product $Q_{q}^{\mathrm{(out)}}Q_{p}^{\mathrm{(out)}}$
of the output modes from the cavity. The spectrum $S_{qp}(\omega)$
is a complex function, so the blue continuous curve is its real part
and the red dashed curve is its imaginary part. The parameters for
the plots are $\gamma=1$, $\Delta=20$, $\kappa=10$, and $F=i3.9$.
\protect\label{cross-quad-spectrum}}
\end{figure}

The Fourier transforms of the intracavity centered quadrature variables
are found as
\begin{equation}
\tilde{Q_{q}}(\omega)=\frac{f_{-}N_{1+}+g_{-}N_{2-}}{g_{+}g_{-}-f_{-}f_{+}}\label{q-trans}
\end{equation}
 and
\begin{equation}
\tilde{Q_{p}}(\omega)=\frac{f_{+}N_{2-}+g_{+}N_{1+}}{g_{+}g_{-}-f_{-}f_{+}},\label{p-trans}
\end{equation}
where we use terms defined by 
\begin{equation}
f_{\pm}(\omega)=-\left(\frac{\gamma}{2}+i\omega\right)\pm6\kappa\langle q\rangle_{sss}\label{f-terms}
\end{equation}
and
\begin{equation}
g_{\pm}=6\kappa\langle p\rangle_{sss}\pm\Delta,\label{g-terms}
\end{equation}
 and the noise functions are written as 
\begin{equation}
N_{1+}=\sqrt{\gamma/4}\tilde{\zeta_{1}}(\omega)+\sqrt[3]{3\kappa/8}\tilde{\xi_{+}}(\omega)\label{n1-terms}
\end{equation}
 and 
\begin{equation}
N_{2-}=\sqrt{\gamma/4}\tilde{\zeta_{2}}(\omega)-i\sqrt[3]{3\kappa/8}\tilde{\xi_{-}}(\omega).\label{n2-terms}
\end{equation}

With all these pieces, we can finally calculate the spectral densities,
including the spectra $S_{q}(\omega)$ and $S_{p}(\omega)$ related
to the quadratures $Q_{q}^{\mathrm{(out)}}$ and $Q_{p}^{\mathrm{(out)}}$.
Both spectra are real functions and are presented in Fig. (\ref{squeezing-spectrum}).
We observe the squeezing phenomenon depending on the frequency relative
to the cavity resonance frequency. Around zero detuning, the quadrature
$Q_{p}^{\mathrm{(out)}}$ exhibits noise squeezing, with its spectrum
dropping below the vacuum level (the standard quantum limit - SQL),
consequently the quadrature $Q_{q}^{\mathrm{(out)}}$ receives excess
noise, remaining above the SQL. However, this scenario is reversed
at higher frequency. After both quadratures exhibit noise peaks, it
is $Q_{q}^{\mathrm{(out)}}$ that has squeezed noise, and $Q_{p}^{\mathrm{(out)}}$
noise raises. This intricate feature is a manifestation of a new type
of squeezing, which is a characteristic trait of the third-order nonlinearity.
For consistency, it is possible to show that the product of $S_{q}(\omega)$
and $S_{p}(\omega)$ obeys the Heisenberg inequality, that is, the
product of both spectra must remain above the minimum value set by
the vacuum state at all frequencies.

The cross spectral density $S_{qp}(\omega)$ is a complex function,
so it must be presented in two plots, one to real part and the other
to imaginary part, as shown in Fig. (\ref{cross-quad-spectrum}).
The interpretation of $S_{qp}(\omega)$ is less clear. This function
is related to covariance between the quadratures. Its value is associated
with the angular position of the quasiprobability density in phase-space,
reflecting the frequency's dependence on the phase shift of the mode's
oscillatory components.

We note in both Figs. (\ref{squeezing-spectrum}) and (\ref{cross-quad-spectrum})
resonant noise peaks at symmetric frequencies. These peaks are results
of the denominators of expressions (\ref{q-trans}) and (\ref{p-trans}),
which are complicated functions of the dissipation factor $\gamma$,
detuning $\Delta$, and the nonlinearity constant $\kappa$. Another
feature of the second-order spectra is the property $S_{ij}(-\omega)=S_{ij}(\omega)^{*}$,
obtained for typical real processes \citep{Gardiner-Zoller,Gardiner}.
This is not the case to the third-order spectra, presented in what
follows. At this point, we should highlight that the degenerate three-photon
down-conversion with classical pump yields a stochastic process in
which the dissipative noise and the third-order nonlinear noise are
explicitly separated in Eqs. (\ref{eq-a}) and (\ref{langevin}).
Hence the second and third-order statistical properties are very distinct
and provide a clear scenario to test experimentally the nonlinear
optical process and our theoretical method.

Since the system is fundamentally non-Gaussian, as is explicit from
the third-order noise in Eq. (\ref{langevin}), we must look for methods
beyond second-order mean values. Inspired by the two-dimensional spectroscopy
in Nuclear Magnetic Resonance (NMR) \citep{Ernst-Bodenhausen-Wokaun},
we propose the introduction of multi-time correlation functions, in
particular third-order three-time correlation functions, for studying
non-Gaussian processes, 
\begin{equation}
\phi_{ijk}(\tau_{1},\tau_{2})=\left\langle Q_{i}^{\mathrm{(out)}}(t+\tau_{1})Q_{j}^{\mathrm{(out)}}(t+\tau_{2})Q_{k}^{\mathrm{(out)}}(t)\right\rangle ,\label{3correlation-quad}
\end{equation}
 whose Fourier transforms define the respective two-dimensional (2D)
spectral densities, 
\begin{equation}
S_{ijk}(\omega_{1},\omega_{2})=\frac{1}{(2\pi)^{2}}\int_{-\infty}^{+\infty}d\tau_{1}e^{-i\omega_{1}\tau_{1}}\int_{-\infty}^{+\infty}d\tau_{2}e^{-i\omega_{2}\tau_{2}}\phi_{ijk}(\tau_{1},\tau_{2}).\label{3spectrum}
\end{equation}
In NMR, the 2D spectra are obtained as responses to pulse sequences
that generally depend on two time parameters. Here, they indeed reflect
the existence of non-null higher-order correlation functions in non-Gaussian
processes, as well as higher-order cumulants must be non-null in such
cases. Similar to the method presented here, a previous study also
utilized multi-time correlations to investigate the spectrum generated
by three-photon parametric down-conversion, but focused on the entanglement
properties of the generated photons \citep{Chekhova05}.

\begin{figure*}
\begin{centering}
\includegraphics[width=0.2\textwidth]{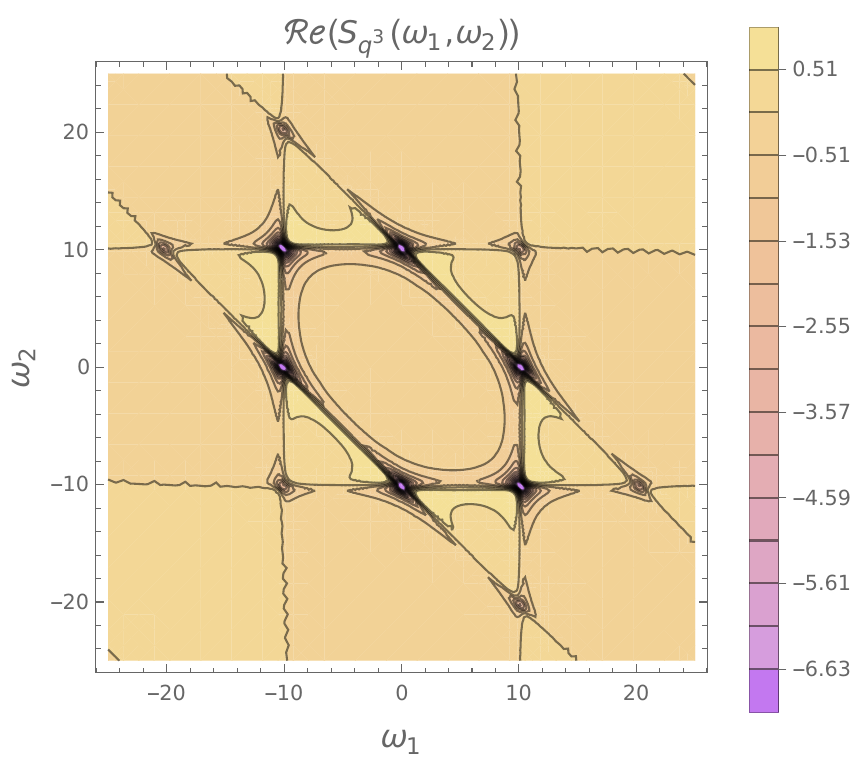}\quad{}\includegraphics[width=0.2\textwidth]{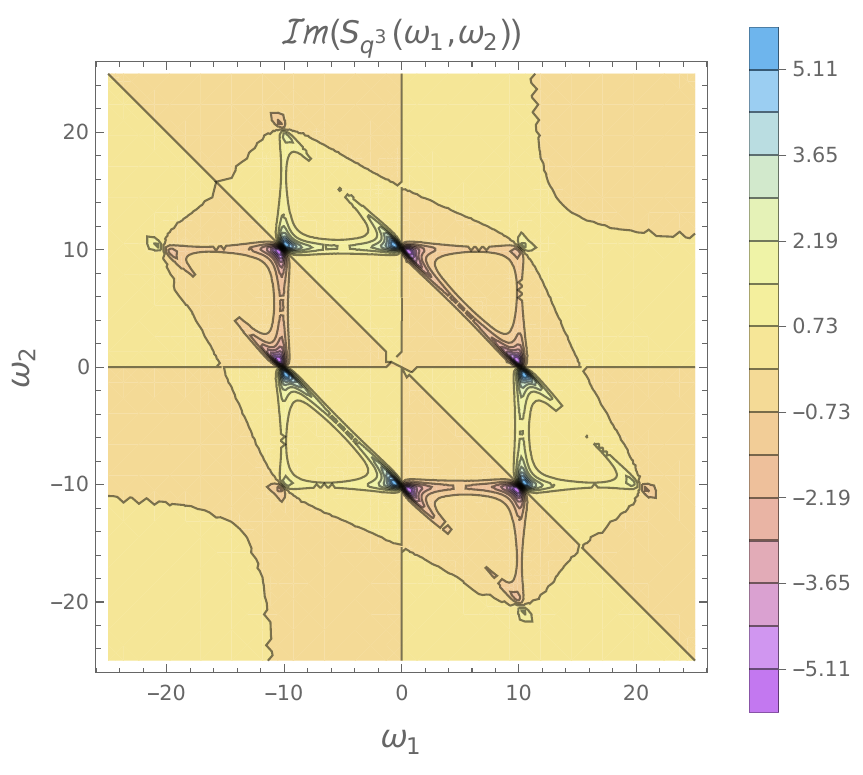}\quad{}\includegraphics[width=0.2\textwidth]{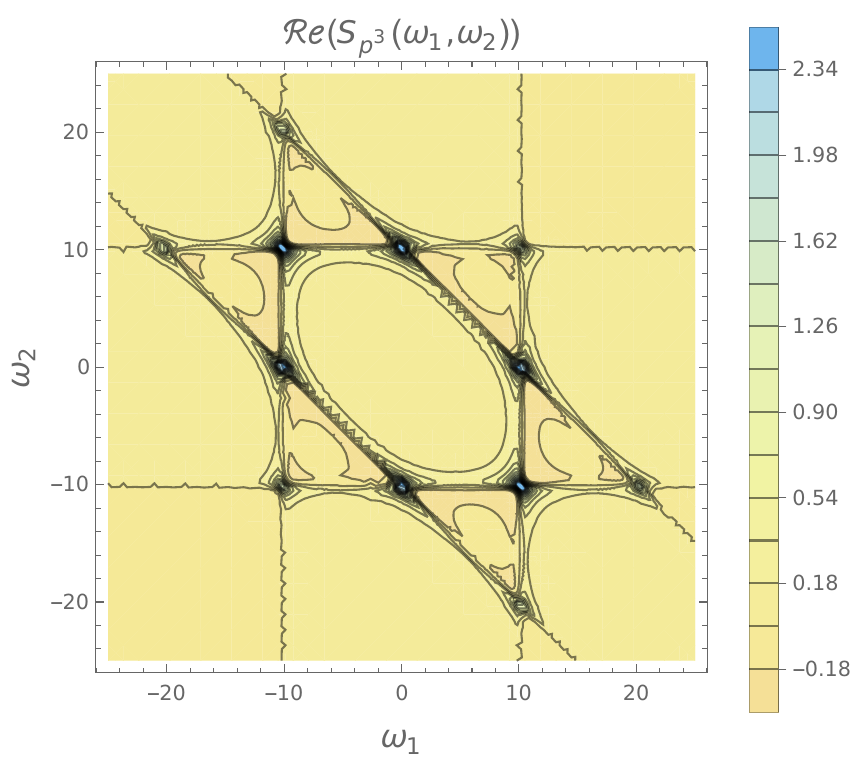}\quad{}\includegraphics[width=0.2\textwidth]{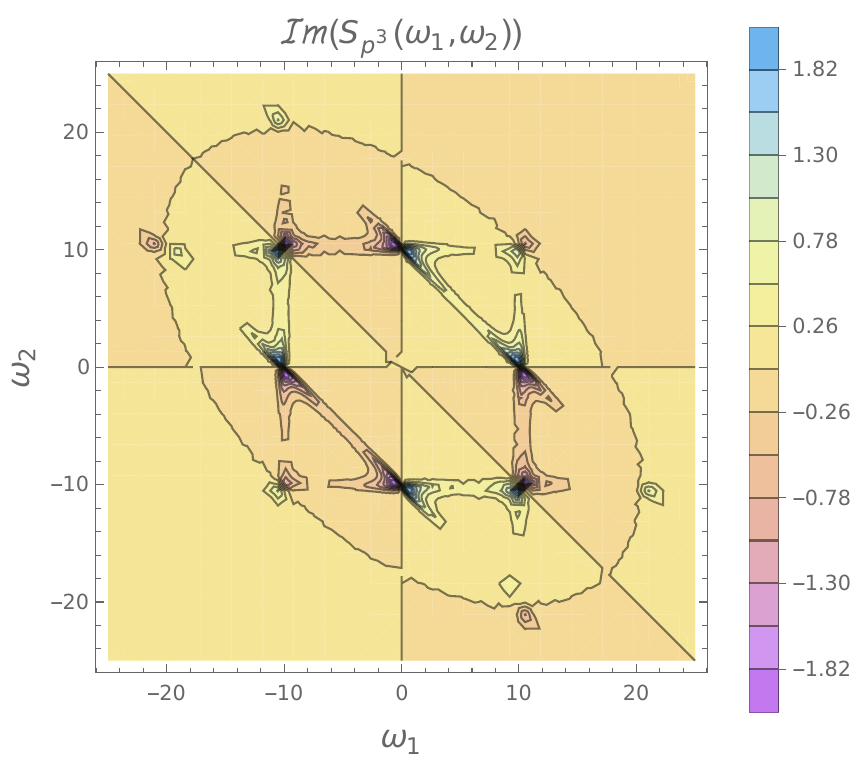}
\par\end{centering}
\caption{2D Noise spectra of the output quadratures $Q_{q}^{\mathrm{(out)}}$
and $Q_{p}^{\mathrm{(out)}}$. The spectra $S_{q^{3}}\equiv S_{qqq}$
and $S_{p^{3}}\equiv S_{ppp}$ are complex functions, so the real
and imaginary parts are shown separately. The parameters for the plots
are $\gamma=1$, $\Delta=20$, $\kappa=10$, and $F=i3.9$. \protect\label{2d-spectrum-direct}}
\end{figure*}

\begin{figure*}
\begin{centering}
\includegraphics[width=0.2\textwidth]{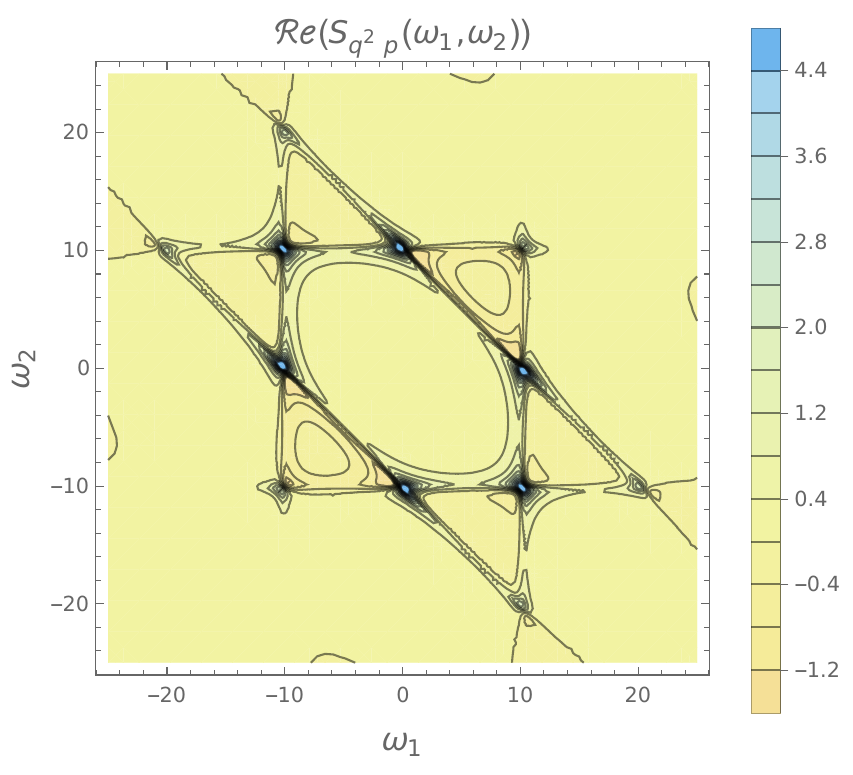}\quad{}\includegraphics[width=0.2\textwidth]{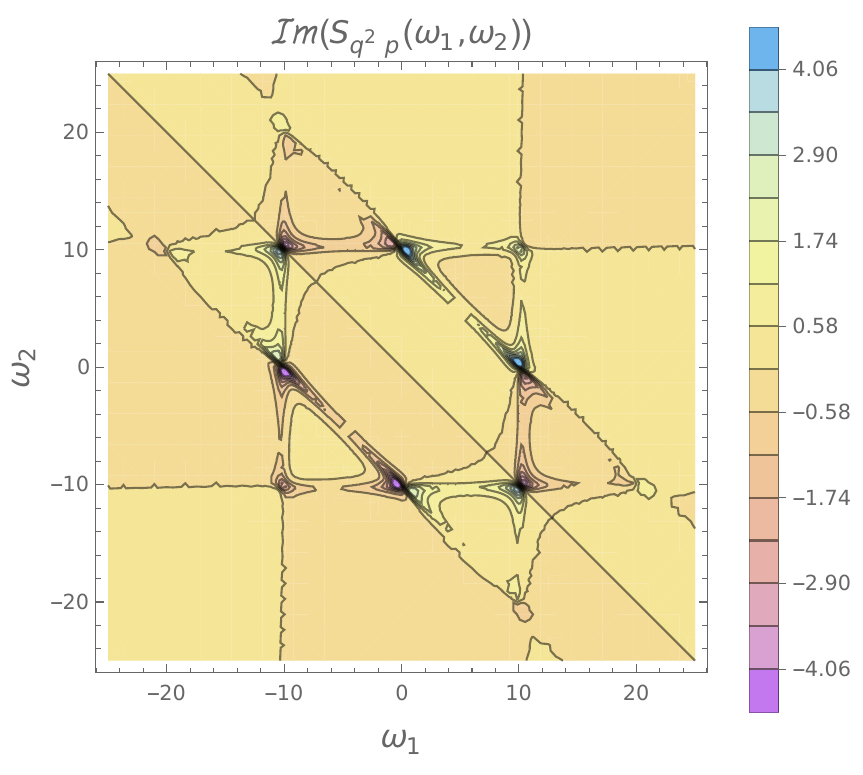}\quad{}\includegraphics[width=0.2\textwidth]{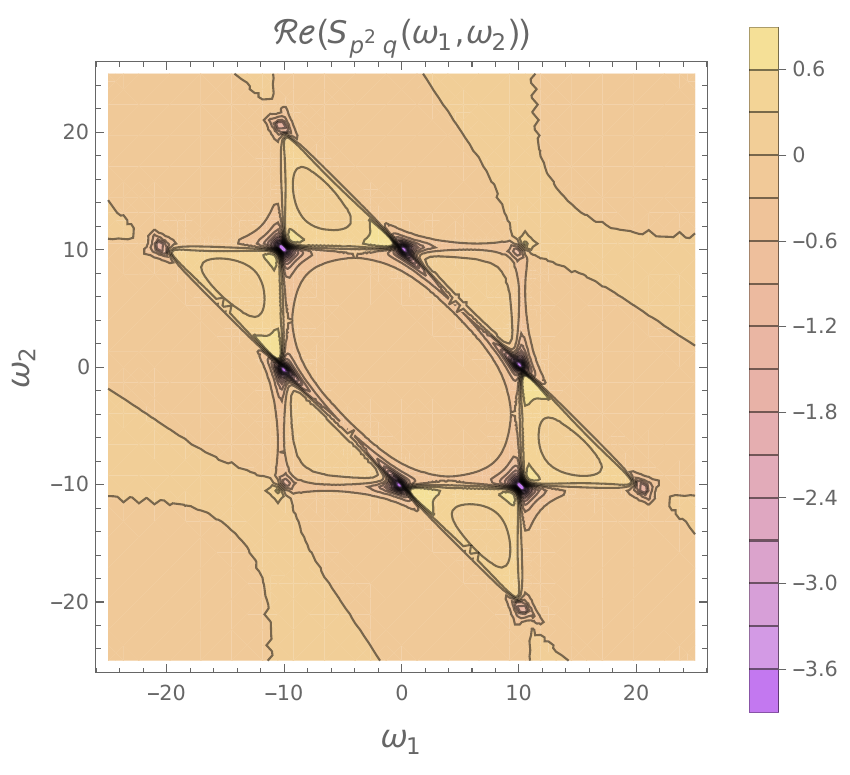}\quad{}\includegraphics[width=0.2\textwidth]{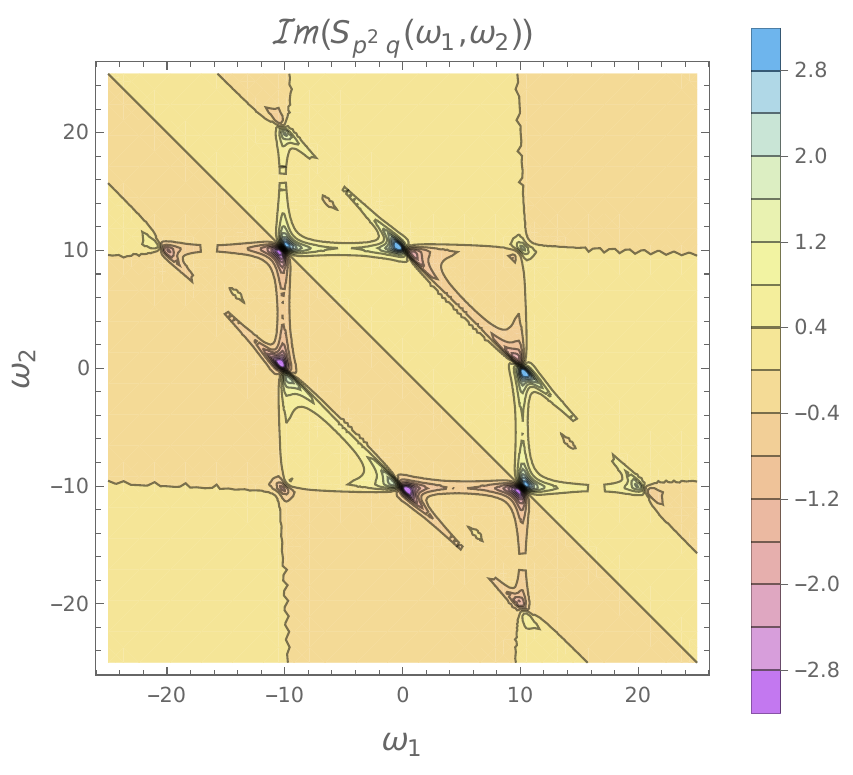}
\par\end{centering}
\caption{2D Cross noise spectra of the combinations of three output quadrature
products for $Q_{q}^{\mathrm{(out)}}$ and $Q_{p}^{\mathrm{(out)}}$.
The spectra $S_{q^{2}p}\equiv S_{qqp}$ and $S_{p^{2}q}\equiv S_{ppq}$
are complex functions, so the real and imaginary parts are shown separately.
The parameters for the plots are $\gamma=1$, $\Delta=20$, $\kappa=10$,
and $F=i3.9$. \protect\label{2d-spectrum-cross}}
\end{figure*}

The 2D spectral densities, $S_{ijk}(\omega_{1},\omega_{2})$, are
related with the third-order correlation function of the Fourier-transformed
quantities by 
\begin{equation}
\left\langle \tilde{Q_{i}}^{\mathrm{(out)}}(\omega_{1})\tilde{Q_{j}}^{\mathrm{(out)}}(\omega_{2})\tilde{Q_{k}}^{\mathrm{(out)}}(-\omega)\right\rangle =S_{ijk}(\omega_{1},\omega_{2})\delta(\omega_{1}+\omega_{2}-\omega)\label{3correlation-fourier}
\end{equation}
 which provides a way to calculate the respective spectra. This approach
is also applied to the Fourier transforms of the third-order noise
correlations (\ref{3noise-langevin-delta}) and (\ref{3noise-langevin-null}).
The nonvanishing correlations are transformed in
\begin{equation}
\left\langle \tilde{\xi}_{+}(\omega)\tilde{\xi}_{+}(\omega^{\prime})\tilde{\xi}_{+}(\omega^{\prime\prime})\right\rangle =\left\langle \tilde{\xi}_{+}(\omega)\tilde{\xi}_{-}(\omega^{\prime})\tilde{\xi}_{-}(\omega^{\prime\prime})\right\rangle =\delta(\omega+\omega^{\prime}+\omega^{\prime\prime}),\label{3noise-fourier}
\end{equation}
and the null correlation cases are trivial. Bringing together the
transformed quadratures (\ref{q-trans}) and (\ref{p-trans}) and
Eqs. (\ref{3correlation-fourier}) and (\ref{3noise-fourier}), we
find the third-order spectra $S_{ijk}(\omega_{1},\omega_{2})$ and
plot the respective graphs, presented in Figs. \ref{2d-spectrum-direct}
and \ref{2d-spectrum-cross}.

As in the magnetic resonance \citep{Ernst-Bodenhausen-Wokaun}, the
functions $S_{ijk}(\omega_{1},\omega_{2})$ are complex and generally
manifest mixed absorptive and dispersive aspects. In the present article,
we find four relevant 2D spectral densities: $S_{qqq}(\omega_{1},\omega_{2})$,
$S_{ppp}(\omega_{1},\omega_{2})$, $S_{qqp}(\omega_{1},\omega_{2})$,
$S_{ppq}(\omega_{1},\omega_{2})$. Other cases are only rotated and
rescaled versions, which do not add information. 

Third-order spectral densities are all complex functions, as they
are Fourier transforms of three-time correlation functions, which
are themselves complex functions, given that there is no direct way
to express them as symmetric products of conjugate variables, unlike
the case for two-time correlation functions. The fact that three-time
correlation functions and their respective 2D spectra are complex-valued
functions is not a novelty, but an inherent feature of this spectroscopic
analysis method \citep{Ernst-Bodenhausen-Wokaun}. The repetitive
structure observed in Figs. \ref{2d-spectrum-direct} and \ref{2d-spectrum-cross}
suggests significant information redundancy in these plots, nevertheless
we present all these compositions for the sake of completeness. As
with 2D NMR, we can analyze the resonance structures based on their
response, whether they are dispersive and/or absorptive, with the
latter aspect understood as the dissipative nature of the cavity.
It is observed that the real parts of the spectra consist of hybrid
responses comprising both dissipative and dispersive components, in
turn the imaginary parts are predominantly dispersive. This dispersive
predominance is characteristic of the nonlinear process, which inherently
involves a phase shift in the system behavior. Furthermore, the frequency
values $\omega_{1}$ and $\omega_{2}$ at which noise peaks appear
also encode information regarding the nonlinear dynamics, as they
are determined by the denominators of Eqs. (\ref{q-trans}) and (\ref{p-trans}),
following the features of the second-order spectral densities. The
triple products of these denominators in Eq. (\ref{3correlation-fourier})
and their off-resonance noise peaks are the reason to the sixfold
repetitive structure in the 2D spectra.

\section{Discussion}

Given the known issues regarding the non-existence of steady states
in a pure generalized squeezing Hamiltonian \citep{Fisher84,Braunstein87,Braunstein90,Ashhab25,Hillery90,Olsen02},
this article has studied a feasible model in optical cavities or in
a circuit QED system for three-photon squeezing, accompanied by dissipative
and driving terms. We should note the differences compared to other
previously studied systems \citep{Felbinger98,Bajer91,Denys21}, which
consider quantized pump accompanied by an applied oscillating term.
The specific choices made here were guided by experimental studies
for which the pumping and driving modes were modeled in a similar
way \citep{Douady04,Bencheikh07,Gravier08,Bertrand25,SandboChang20}.
Furthermore, we have applied our study to stationary fields generated
within cavities (whether optical or superconducting circuit-based),
which also distinguishes this work from studies focused on traveling-wave
regimes \citep{Douady04,Bertrand25,Dot12,Borshchevskaya15,Okoth19,Corona11a,Corona11b,Cavanna16,Cavanna20,Banic22,Bacaoco25}.
Thus, we believe we have addressed a gap in the study of third-order
nonlinear systems.

Using a mean-field approximation, we found that the system does have
a single stable steady state branch restricted to a limited region
of its parameters. Outside this region, the system does not have stable
states, basically exhibiting an evolution towards divergent values
of the observable quantities. The stability region expands as the
dissipation factor $\gamma$ and the detuning $\Delta$ increase,
shrinking to a region of zero measure for an undamped system at resonance
($\gamma=0$ and $\Delta=0$). On the other hand, an inverse relationship
holds for nonlinearity factor $\kappa$. This is consistent with previously
established results, which showed that there are no steady solutions
for a generalized parametric down-conversion Hamiltonian \citep{Fisher84,Braunstein87,Braunstein90,Hillery90,Olsen02}.
However, we have shown that even with classical pump, but with dissipative
and driving terms, we can obtain finite stable steady results.

We have also considered third-order noise originated from the Kramers-Moyal
equation of the Wigner function. Although there are difficulties in
interpreting the third-order or higher partial differential equations
for quasiprobabilities and converting them to stochastic equations,
we applied a method that extends the domain of Wigner function, so
that it is possible interpret it as a stochastic process \citep{Drummond14,Drummond17}.
We have considered that this positive-Wigner function method deserves
more attention, because it offers a new way to study stochastic quantum
systems via Wigner functions, potentially providing a more accessible
physical interpretation. Furthermore, it serves as an additional computational
tool for studying nonlinear systems, as demonstrated by the decoupling
of calculations regarding dissipation and nonlinear noises.

Applying the stochastic differential equations of the system to the
input-output field formalism \citep{Gardiner-Zoller,Collett84,Gardiner85},
we have obtained a system of equations that has been linearized. From
this, we perform a Fourier analysis of the output modes from the nonlinear
cavity. Studying the output quadrature spectra, we were able to verify
the existence of quantum noise squeezing with unique features distinct
from ordinary squeezing. Unlike second-order squeezing, in three-photon
squeezing we observe two symmetric noise peaks away from cavity resonance
($\Delta=0$), which separate two frequency regimes with different
behaviors: for smaller detunings, one quadrature is squeezed; whereas
for larger detunings the squeezed quadrature is inverted. It is also
observed that the noise level reduction is limited and less pronounced
in three-photon squeezing. This is due to the triply symmetric geometry
of squeezing in phase-space and the squeezing limitation imposed by
the Heisenberg inequality. It is also noted that the squeezing effect
is greater near the stability boundary of the steady states, at the
limit of the driving mode amplitude near the bifurcation, with squeezing
being zero when the driving mode is null.

In addition to the spectral analysis associated with the two-time
correlation functions, we were able to calculate 2D spectral densities,
so that it was possible to observe the dispersive and dissipative
aspects and the resonant noise peaks off the cavity resonance of the
three-photon squeezing. We have followed the mathematical method developed
for magnetic resonance spectroscopy in two frequency dimensions \citep{Ernst-Bodenhausen-Wokaun}.
Since the 2D spectra are Fourier transforms of the three-time correlation
functions of the quadratures centered in their stable state, they
are explicitly non-Gaussian, because the third-order cumulants are
null for Gaussian systems. Hence we have shown a novel analysis for
non-Gaussian and nonlinear systems.

Through our study of the driven dissipative three-photon down-conversion,
we were able to deepen our understanding of the third-order squeezing
operation, which is a potential process for use in continuous-variable
quantum information and computing protocols \citep{Minganti23,Labay-Mora23,Labay-Mora24}.
Here, we provide further insights for implementing such nonlinear
squeezing operations in fault-tolerant computing processes or for
constructing a complete set of logic gates \citep{GKP01,Zheng21,Hahn22}.
Although we focused on the degenerate down-conversion case, this study
serves as a starting point for enhancing our understanding of the
nondegenerate process. We understand the application of this analysis
to the generation of tripartite entangled states as a natural next
step. Despite the entanglement has already been the subject of numerous
studies \citep{Chekhova05,Agusti20,Zhang21,Tian22,Zhang23,Wei24},
our work offers additional insight into stability and spectral analysis.
\begin{acknowledgments}
We would like to thank Dr. Rodrigo Cuzinatto for his help at crucial
points during the development of this article.
\end{acknowledgments}


\begin{thebibliography}{99}
\bibitem[(2004)]{Gardiner-Zoller} C. W. Gardiner and P. Zoller, Quantum
Noise: A Handbook of Markovian and Non-Markovian Quantum Stochastic
Methods with Applications to Quantum Optics, 3rd Ed. (Springer-Verlag,
Berlin, Heidelberg, 2004).

\bibitem[(1997)]{Scully-Zubairy} M. O. Scully and M. S. Zubairy,
\textit{Quantum Optics} (Cambridge University, 1997).

\bibitem[(2021)]{cQED21} A. Blais, A. L. Grimsmo, S. M. Girvin, and
A. Wallraff, Circuit quantum electrodynamics, Rev. Mod. Phys. 93,
025005 (2021).

\bibitem[(2001)]{GKP01} D. Gottesman, A. Kitaev, and J. Preskill,
Encoding a qubit in an oscillator, Phys. Rev. A 64, 012310 (2001).

\bibitem[(2021)]{Zheng21} Y. Zheng, O. Hahn, P. Stadler, P. Holmvall,
F. Quijandría, A. Ferraro, and G. Ferrini, Gaussian Conversion Protocols
for Cubic Phase State Generation, PRX Quantum 2, 010327 (2021).

\bibitem[(2022)]{Hahn22} O. Hahn, P. Holmvall, P. Stadler, G. Ferrini,
and A. Ferraro, Deterministic Gaussian conversion protocols for non-Gaussian
single-mode resources, Phys. Rev. A 105, 062446 (2022).

\bibitem[(1984)]{Fisher84} R. A. Fisher, M. M. Nieto, and V. D. Sandberg,
Impossibility of naively generalizing squeezed coherent states, Phys.
Rev. D 29, 1107 (1984).

\bibitem[(1987)]{Braunstein87} S. L. Braunstein and R. I. McLachlan,
Generalized squeezing, Phys. Rev. A 35, 1659 (1987).

\bibitem[(1990)]{Braunstein90} S. L. Braunstein and C. M. Caves,
Phase and homodyne statistics of generalized squeezed states, Phys.
Rev. A 42, 4115 (1990).

\bibitem[(2025)]{Ashhab25} S. Ashhab and M. Ayyash, Properties and
dynamics of generalized squeezed states, New J. Phys. 27, 054104 (2025).

\bibitem[(1990)]{Hillery90} M. Hillery, Photon number divergence
in the quantum theory of n-photon down conversion, Phys. Rev. A 42,
498 (1990).

\bibitem[(1992)]{Drobny92} G. Drobný and I. Jex, Statistics of field
modes in the process of k-photon down-conversion with a quantized
pump, Phys. Rev. A 45, 4897 (1992).

\bibitem[(1992)]{Tanas92} R. Tanaś and Ts. Gantsog, Phase properties
of fields generated in a multiphoton down-converter, Phys. Rev. A
45, 5031 (1992).

\bibitem[(1997)]{Banaszek97} K. Banaszek and P. L. Knight, Quantum
interference in three-photon down-conversion, Phys. Rev. A 55, 2368
(1997).

\bibitem[(2002)]{Olsen02} M. K. Olsen, L. I. Plimak, and M. Fleischhauer,
Quantum-theoretical treatments of three-photon processes, Phys. Rev.
A 65, 053806 (2002).

\bibitem[(1998)]{Felbinger98} T. Felbinger, S. Schiller, and J. Mlynek,
Oscillation and Generation of Nonclassical States in Three-Photon
Down-Conversion, Phys. Rev. Lett. 80, 492 (1998).

\bibitem[(1991)]{Bajer91} J. Bajer, Photon Statistics of the Nth
Subharmonics, J. Mod. Opt. 38, 1085 (1991).

\bibitem[(2021)]{Denys21} M. D. E. Denys, M. K. Olsen, L. S. Trainor,
H. G. L. Schwefel, and A. S. Bradley, Steady states, squeezing, and
entanglement in intracavity triplet down conversion, Opt. Commun.
484, 126699 (2021).

\bibitem[(2004)]{Douady04} J. Douady and B. Boulanger, Experimental
demonstration of a pure third-order optical parametric downconversion
process, Opt. Lett. 29, 2794 (2004).

\bibitem[(2007)]{Bencheikh07} K. Bencheikh, F. Gravier, J. Douady,
A. Levenson, and B. Boulanger, Triple photons: a challenge in nonlinear
and quantum optics, C. R. Physique 8, 206 (2007).

\bibitem[(2008)]{Gravier08} F. Gravier and B. Boulanger, Triple-photon
generation: comparison between theory and experiment, J. Opt. Soc.
Am. B 25, 98 (2008).

\bibitem[(2025)]{Bertrand25} J. Bertrand, V. Boutou, C. Felix, D.
Jegouso, and B. Boulanger, Experimental demonstration and modeling
of near-infrared nonlinear third-order triple-photon generation stimulated
over one mode, APL Quantum 2, 026114 (2025).

\bibitem[(2012)]{Dot12} A. Dot, A. Borne, B. Boulanger, K. Bencheikh,
and J. A. Levenson, Quantum theory analysis of triple photons generated
by a $\chi^{(3)}$ process, Phys. Rev. A 85, 023809 (2012).

\bibitem[(2015)]{Borshchevskaya15} N. A. Borshchevskaya, K. G. Katamadze,
S. P. Kulik, and M. V. Fedorov, Three-photon generation by means of
third-order spontaneous parametric down-conversion in bulk crystals,
Laser Phys. Lett. 12, 115404 (2015).

\bibitem[(2019)]{Okoth19} C. Okoth, A. Cavanna, N. Y. Joly, and M.
V. Chekhova, Seeded and unseeded high-order parametric down-conversion,
Phys. Rev. A 99, 043809 (2019).

\bibitem[(2022)]{Bencheikh22} K. Bencheikh, M. F. B. Cenni, E. Oudot,
V. Boutou, C. Félix, J. C. Prades, A. Vernay, J. Bertrand, F. Bassignot,
M. Chauvet, F. Bussières, H. Zbinden, A. Levenson, and B. Boulanger,
Demonstrating quantum properties of triple photons generated by $\chi^{3}$
processes, Eur. Phys. J. D 76, 186 (2022).

\bibitem[(2020)]{SandboChang20} C. W. Sandbo Chang, C. Sabín, P.
Forn-Díaz, F. Quijandría, A. M. Vadiraj, I. Nsanzineza, G. Johansson,
and C. M. Wilson, Observation of Three-Photon Spontaneous Parametric
Down-Conversion in a Superconducting Parametric Cavity, Phys. Rev.
X 10, 011011 (2020).

\bibitem[(2011)]{Corona11a} M. Corona, K. Garay-Palmett, and A. B.
U’Ren, Experimental proposal for the generation of entangled photon
triplets by third-order spontaneous parametric downconversion in optical
fibers, Opt. Lett. 36, 190 (2011).

\bibitem[(2011)]{Corona11b} M. Corona, K. Garay-Palmett, and A. B.
U’Ren, Third-order spontaneous parametric down-conversion in thin
optical fibers as a photon-triplet source, Phys. Rev. A 84, 033823
(2011).

\bibitem[(2016)]{Akbari16} M. Akbari and A. A. Kalachev, Third-order
spontaneous parametric down-conversion in a ring microcavity, Laser
Phys. Lett. 13, 115204 (2016).

\bibitem[(2016)]{Cavanna16} A. Cavanna, F. Just, X. Jiang, G. Leuchs,
M. V. Chekhova, P. St.J. Russell, and N. Y. Joly, Hybrid photonic-crystal
fiber for single-mode phase matched generation of third harmonic and
photon triplets, Optica 3, 952 (2016). 

\bibitem[(2020)]{Cavanna20} A. Cavanna, J. Hammer, C. Okoth, E. Ortiz-Ricardo,
H. Cruz-Ramirez, K. Garay-Palmett, A. B. U’Ren, M. H. Frosz, X. Jiang,
N. Y. Joly, and M. V. Chekhova, Progress toward third-order parametric
down-conversion in optical fibers, Phys. Rev. A 101, 033840 (2020).

\bibitem[(2022)]{Banic22} M. Banic, M. Liscidini, and J. E. Sipe,
Resonant and nonresonant integrated third-order parametric down-conversion,
Phys. Rev. A 106, 013710 (2022).

\bibitem[(2025)]{Bacaoco25} M. Y. Bacaoco, K. Koshelev, and A. S.
Solntsev, Third-Order Spontaneous Parametric Down Conversion in Dielectric
Nonlinear Resonant Metasurfaces, ACS Photonics 12, 4397 (2025).

\bibitem[(2014)]{Drummond14} P. D. Drummond, Fundamentals of higher
order stochastic equations, J. Phys. A: Math. Theor. 47, 335001 (2014).

\bibitem[(2017)]{Drummond17} P. D. Drummond, Higher-order stochastic
differential equations and the positive Wigner function, Phys. Rev.
A 96, 062104 (2017).

\bibitem[(1997)]{Ernst-Bodenhausen-Wokaun} R. R. Ernst, G. Bodenhausen,
and A. Wokaun, Principles of Nuclear Magnetic Resonance in One and
Two Dimensions (Oxford University Press, Oxford, 1997).

\bibitem[(1984)]{Collett84} M. J. Collett and C. W. Gardiner, Squeezing
of intracavity and traveling-wave light fields produced in parametric
amplification, Phys. Rev. A 30, 1386 (1984).

\bibitem[(1985)]{Gardiner85} C. W. Gardiner and M. J. Collett, Input
and output in damped quantum systems: Quantum stochastic differential
equations and the master equation, Phys. Rev. A 31, 3761 (1985).

\bibitem[(1967)]{Pawula67} R. Pawula, Generalizations and extensions
of the Fokker- Planck-Kolmogorov equations, IEEE Trans. Inf. Theory
13, 33 (1967).

\bibitem[(1980)]{Drummond80a} P. D. Drummond and D. F. Walls, Generalised
P-representations in quantum optics, J. Phys. A: Math. Gen. 13, 2353
(1980).

\bibitem[(2004)]{Gardiner} C. W. Gardiner, Handbook of Stochastic
Methods: for Physics, Chemistry and the Natural Sciences, 3rd Ed.
(Springer-Verlag, Berlin, Heidelberg, 2004).

\bibitem[(2018)]{Strogatz} S. H. Strogatz, Nonlinear Dynamics and
Chaos: With Applications to Physics, Biology, Chemistry, and Engineering,
2nd Ed. (CRC Press, Boca Raton, London, New York, 2018).

\bibitem[(1977)]{Gardiner77} C. W. Gardiner and S. Chaturvedi, The
Poisson Representation. I. A New Technique for Chemical Master Equations,
J. Stat. Phys. 17, 429 (1977).

\bibitem[(2020)]{Chekhova05} M. V. Chekhova, O. A. Ivanova, V. Berardi,
and A. Garuccio, Spectral properties of three-photon entangled states
generated via three-photon parametric down-conversion in a $\chi^{(3)}$
medium, Phys. Rev. A 72, 023818 (2005).

\bibitem[(2023)]{Minganti23} F. Minganti, V. Savona, and A. Biella,
Dissipative phase transitions in n-photon driven quantum nonlinear
resonators, Quantum 7, 1170 (2023).

\bibitem[(2023)]{Labay-Mora23} A. Labay-Mora, R. Zambrini, and G.
L. Giorgi, Quantum Associative Memory with a Single Driven-Dissipative
Nonlinear Oscillator, Phys. Rev. Lett 130, 190602 (2023).

\bibitem[(2024)]{Labay-Mora24} A. Labay-Mora, R. Zambrini, and G.
L. Giorgi, Quantum memories for squeezed and coherent superpositions
in a driven-dissipative nonlinear oscillator, Phys. Rev. A 109, 032407
(2024).

\bibitem[(2020)]{Agusti20} A. Agustí, C. W. Sandbo Chang, F. Quijandría,
G. Johansson, C. M. Wilson, and C. Sabín, Tripartite Genuine Non-Gaussian
Entanglement in Three-Mode Spontaneous Parametric Down-Conversion,
Phys. Rev. Lett. 125, 020502 (2020).

\bibitem[(2021)]{Zhang21} D. Zhang, Y. Cai, Z. Zheng, D. Barral,
Y. Zhang, M. Xiao, and K. Bencheikh, Non-Gaussian nature and entanglement
of spontaneous parametric nondegenerate triple-photon generation,
Phys. Rev. A 103, 013704 (2021).

\bibitem[(2022)]{Tian22} M. S. Tian, Y. Xiang, F. X. Sun, M. Fadel,
and Q. Y. He, Characterizing multipartite non-Gaussian entanglement
for a three-mode spontaneous parametric down-conversion process, Phys.
Rev. Appl. 18, 024065 (2022).

\bibitem[(2023)]{Zhang23} D. Zhang, D. Barral, Y. Zhang, M. Xiao,
and K. Bencheikh, Genuine tripartite Non-Gaussian entanglement, Phys.
Rev. Lett. 130, 093602 (2023).

\bibitem[(2024)]{Wei24} Miaomiao Wei and Huatang Tan, Steady-state
tripartite non-Gaussian entanglement and steering in the output field
from intracavity triple-photon parametric down-conversion, Phys. Rev.
A 110, 023729 (2024).

\end{thebibliography}
\end{document}